\documentclass[reqno,12pt,letterpaper]{amsart}
\usepackage{amsmath,amssymb,amsthm,graphicx,mathrsfs,url}
\usepackage[usenames,dvipsnames]{color}
\usepackage[colorlinks=true,linkcolor=Red,citecolor=Green]{hyperref}
\usepackage{amsxtra}
\usepackage{wasysym} 
\usepackage{graphicx}
\usepackage{subcaption}

\usepackage{graphicx,color}

\def\arXiv#1{\href{http://arxiv.org/abs/#1}{arXiv:#1}}
\usepackage{array}
\newcolumntype{P}[1]{>{\centering\arraybackslash}m{#1}}

\def\wrtext#1{\relax\ifmmode{\leavevmode\hbox{#1}}\else{#1}\fi}

\def\abs#1{\left|#1\right|}

\def\?[#1]{\textbf{[#1]}\marginpar{\Large{\textbf{??}}}}

\let\epsilon=\varepsilon 

\newcommand{\RR}{{\mathbb R}}

\def\norm#1{||\,#1\,||}

\newtheorem{theo}{Theorem}
\newtheorem{prop}{Proposition}[section]

\newtheorem{lemm}[prop]{Lemma}

\numberwithin{equation}{section}

\DeclareMathOperator{\Spec}{Spec}

\DeclareMathOperator{\Hom}{Hom}

\let\Im=\Imag

\let\Re=\Real

\DeclareMathOperator{\supp}{supp}

\DeclareMathOperator{\WF}{WF}

\DeclareMathOperator{\neigh}{neigh}

\def\WFh{\WF_h}

\usepackage{scalerel}

\newcommand\reallywidehat[1]{\arraycolsep=0pt\relax%
\begin{array}{c}
\stretchto{
  \scaleto{
    \scalerel*[\widthof{\ensuremath{#1}}]{\kern-.5pt\bigwedge\kern-.5pt}
    {\rule[-\textheight/2]{1ex}{\textheight}} 
  }{\textheight} %
}{0.5ex}\\           
#1\\                 
\rule{-1ex}{0ex}
\end{array}
}

\def\red#1{\textcolor{red}{#1}}

\begin{document}


\title[Classically forbidden regions in the chiral model of TBG]{Classically forbidden regions in the chiral model of twisted bilayer graphene}

\author{Michael Hitrik}
\address{Department of Mathematics, University of California,
Los Angeles, CA 90095, USA.}
\email{hitrik@math.ucla.edu}

\address{Department of Mathematics, University of California,
Berkeley, CA 94720, USA.}
\email{tzk320581@berkeley.edu}

\author{Maciej Zworski}
\address{Department of Mathematics, University of California,
Berkeley, CA 94720, USA.}
\email{zworski@math.berkeley.edu}


\maketitle

\vspace{-0.2in}

\begin{center}
{\sc With an appendix by Zhongkai Tao and Maciej Zworski}
\end{center}

\vspace{-0.1in}

\begin{abstract}
We establish exponential decay, as the angle of twisting goes to $ 0$, of eigenstates
in a model  of
twisted bilayer graphene (TBG) \cite{magic}, near the hexagon
connecting stacking points. That is done by adapting
microlocal methods \cite{kaka,sam,HiS} used to establish
analytic hypoellipticity \cite{tre,him}. We also discuss
numerical evidence of exponential decay near the center of the hexagon, and
analytic complications involved in establishing that decay.
\end{abstract}

\section{Introduction}
\label{s:int}

Twisted bilayer graphene (TBG) is described by the Bistritzer--MacDonald Hamiltonian \cite{BM11} which was used to predict existence of flat bands and special properties at
a magical angle of twisting of TBG \cite{Cao} -- see
\cite{CGG} and \cite{Wa22} for mathematical derivations of the model. Its chiral limit  is obtained by removing certain
tunneling interactions and it was
very successfully analysed by
Tarnopolsky--Kruchkov--Vishwanath \cite{magic} who explained a mechanism
for the existence of perfectly flat bands.

In coordinates used in
\cite{bhz2} the Hamiltonian is given by
\begin{equation}
\label{eq:Ham}
\begin{gathered}
H ( \alpha ) := \begin{pmatrix} 0 & D ( \alpha )^* \\
D ( \alpha ) & 0 \end{pmatrix} , \ \ \
D ( \alpha ) := \begin{pmatrix} 2 D_{ \bar z } & \alpha U ( z ) \\
\alpha U ( - z ) & 2 D_{\bar z } \end{pmatrix} , \\  2D_{\bar z } = \tfrac 1 i ( \partial_{{x_1}} + i \partial_{x_2} ) , \ \
z = {x_1} + i x_2 \in \mathbb C ,
\end{gathered}
\end{equation}
where $ U ( z ) $ is the Bistritzer--MacDonald potential,
\begin{equation}
  \label{eq:defU} U ( z ) =  - \tfrac{4} 3 \pi i \sum_{ \ell = 0 }^2 \omega^\ell e^{ i \langle z , \omega^\ell K \rangle }, \ \ \ K = \tfrac43 \pi  , \ \  \omega := e^{ 2 \pi i /3 } , \ \ \langle z, w \rangle := \Re ( z \bar w ).
  \end{equation}
The coupling constant $ \alpha $ is a dimensionless parameter which, after suitable rescaling,
corresponds to $ \theta  \simeq 1/\alpha $, the angle of twisting of the two sheets.

The potential $ U $ is periodic with respect to the lattice $ \Gamma$,
\begin{equation}
\label{eq:defG}  \Gamma := 3 \Lambda  , \ \ \ \ \Lambda := \omega \mathbb Z \oplus \mathbb Z , \ \ \ \
\Lambda^* = \frac {4 \pi i}  {\sqrt 3}  \Lambda ,
\end{equation}
with finer periodicity properties with respect to $ \Lambda $ (here $ \Lambda^* $ is the dual lattice to
$ \Lambda$). The remarkable property of $ H( \alpha ) $ is the existence of perfectly flat
bands at the $ 0 $ energy. The $ \alpha $'s for which flat bands occur are known as {\em magical} -- see Becker et al \cite{beta} for a mathematical presentation,
Watson--Luskin \cite{lawa} for the existence of the first real magic $ \alpha $, and
\cite{bhz1} for a different proof which also establishes its simplicity. As  emphasised   in \cite{beta} (see
\S \ref{s:flat} for a brief review)
having a flat band is equivalent to
\begin{equation}
\label{eq:aflat}     \Spec_{ L^2 ( \mathbb C/\Gamma ; \mathbb C^2 )} D ( \alpha) = \mathbb C .
\end{equation}
It is then interesting to understand the 
 structure of the corresponding eigenstates, as well as those in  the {\em protected} two dimensional kernel of $ D ( \alpha ) $ on $ H^1 ( \mathbb C/\Gamma ) $ - see \cite{magic}, \cite[Theorem 1]{beta}
 and \S \ref{s:flat}.

\begin{center}
\begin{figure}
\includegraphics[width=5.75cm]{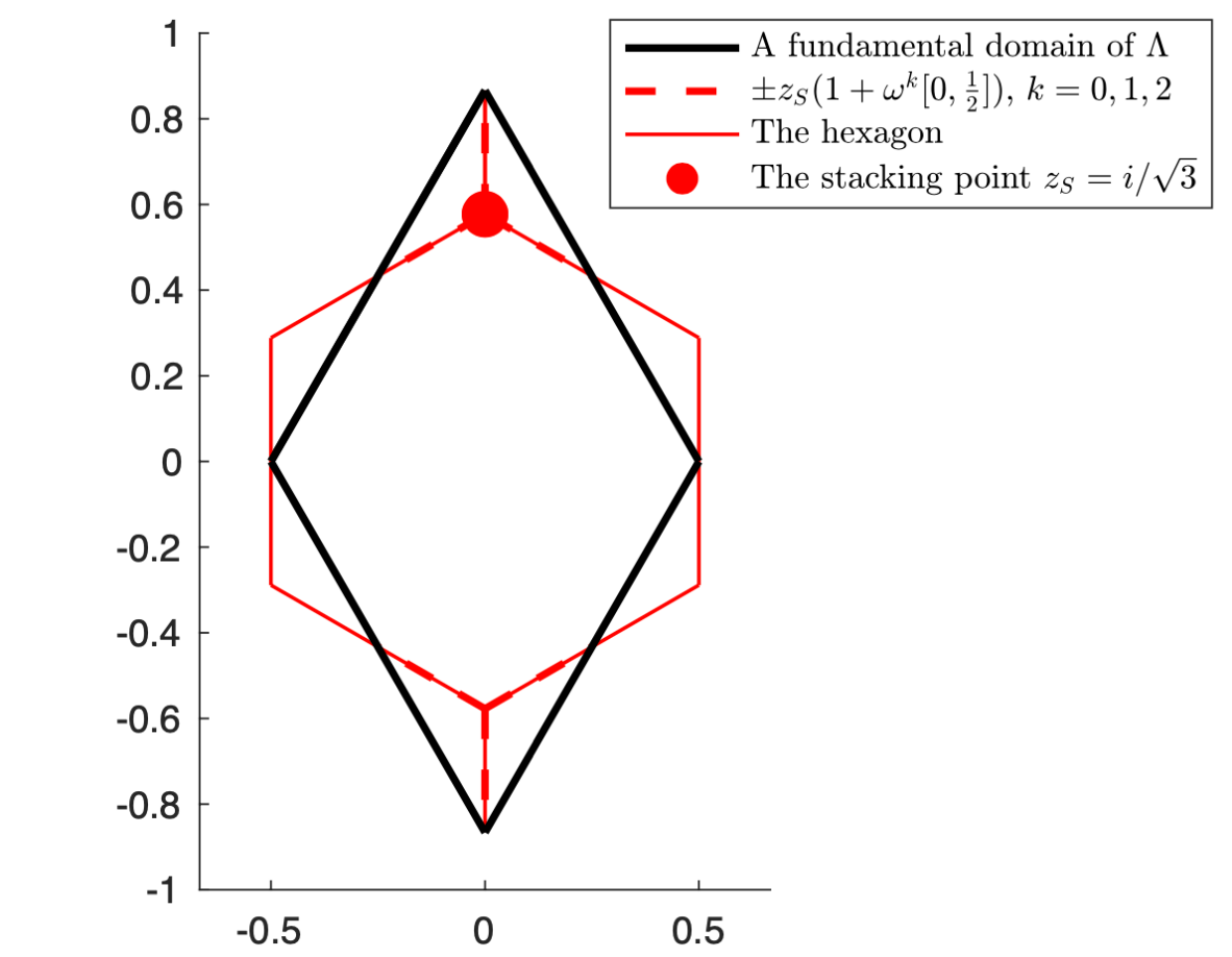} \includegraphics[width=4.1cm]{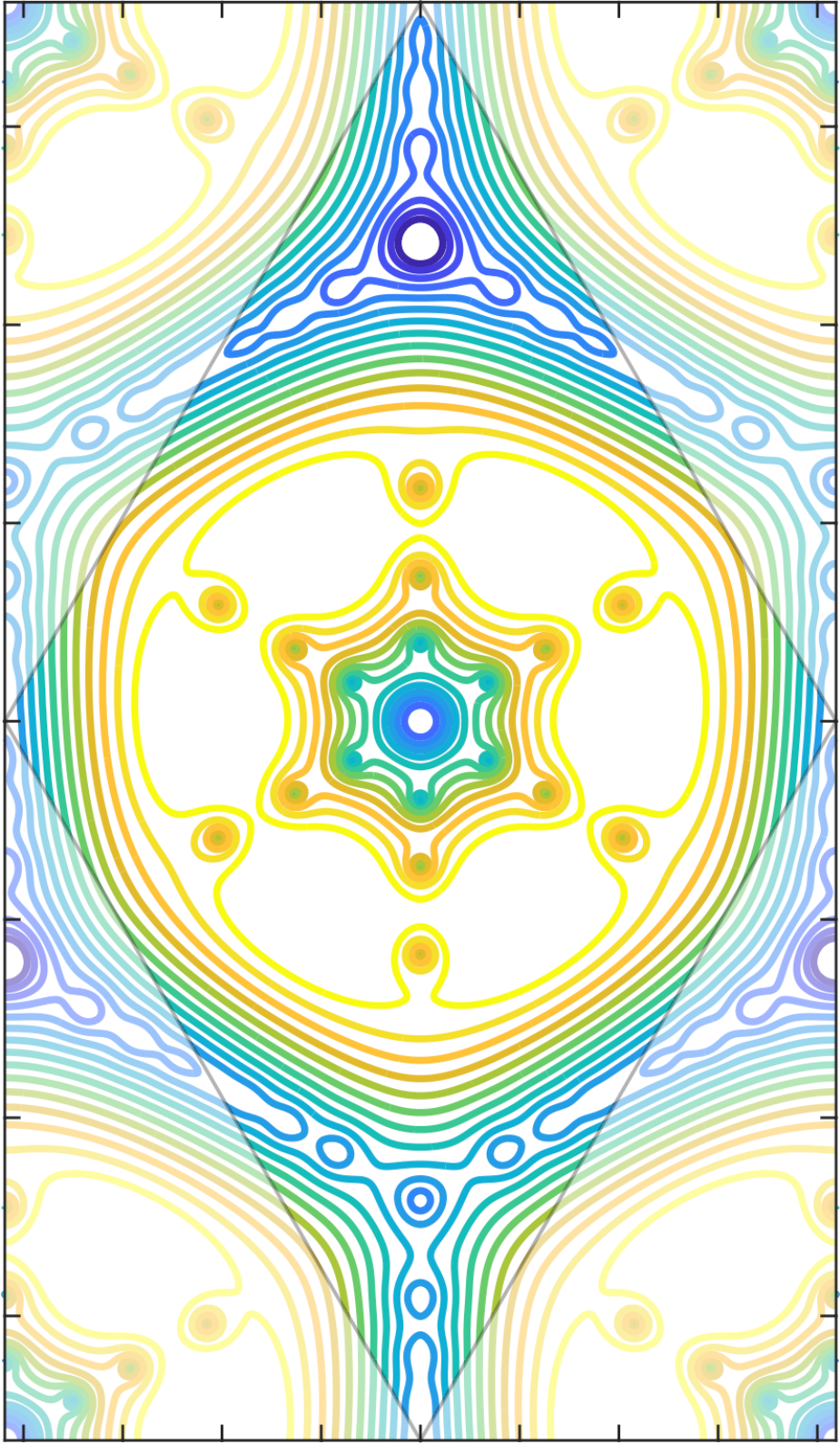} \includegraphics[width=4.65cm]{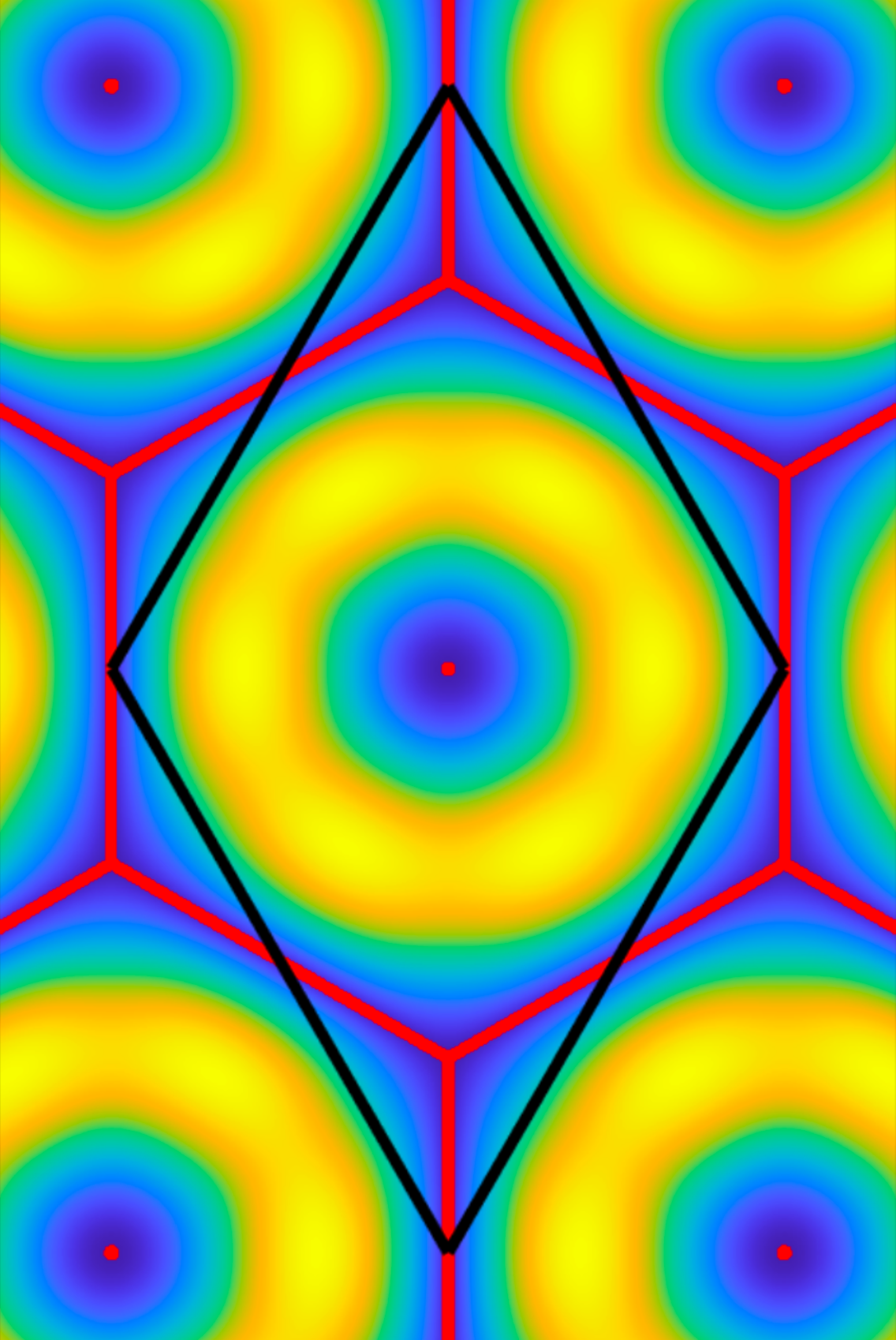}
\caption{Left: the vertices of the hexagon in a fundamental domain of $ \Lambda $
are given by the
{\em stacking points} $ \pm z_S $, $ z_S = i/\sqrt 3 $. They are non-zero
points of high symmetry in the sense that $ \pm \omega z_S \equiv \pm z_S \!\! \mod  \Lambda$.
Center: plot of $ \log |u_K ( \alpha )| $ where $ u_K $ is the protected state
in the kernel of $ D ( \alpha ) $ on $ H^1 ( \mathbb C/\Gamma )$ and $ \alpha = 11.345 $.
Dark blue corresponds to $ |u_K | \simeq 10^{-7} $ and yellow to $ |u_K| \simeq 1 $: we see
exponential decay $ e^{ -c_0 \alpha } $ near the hexagon and near its center. (This figure is borrowed from \cite{beta}.)  Right:  the contour plot of $ |\{ q, \bar q \}|_{q^{-1} ( 0 ) } |$ for $ q $ given by the determinant of the
semiclassical symbol of $ D ( \alpha ) $ (see \eqref{eq:Ham} and \eqref{eq:defU}), $ \alpha = 1/h $; the set
where $ \{ q, \bar q \}|_{q^{-1} ( 0 ) } = 0 $ is in \red{red}. We should stress that the structure of that
set becomes more complicated for other potentials $ U $ satisfying the required symmetries - see
\S \ref{s:flat} and
 \cite[Figure 6]{beta}.
\label{f:1}}
\end{figure}
\end{center}

   The natural asymptotic parameter is $ \alpha \to \infty $,
 which corresponds to the small twisting angle limit (the angle is essentially given by $ 1/\alpha $).
 The operator $ D ( \alpha ) $ is then a semiclassical operator with $ h = 1/\alpha \to 0 $ playing
 the role of the semiclassical parameter.

 A fascinating numerical observation about the asymptotic behaviour of real magic $ \alpha$'s
 was made in \cite{magic}: if $ \alpha_1 < \alpha_2 < \cdots \alpha_j < \cdots $ is the
 sequence of all real $ \alpha$'s for which \eqref{eq:aflat} holds, then
 \begin{equation}
 \label{eq:quant}  \alpha_{j+1} - \alpha_j \simeq \tfrac32 .
 \end{equation}
 This was based on a rough computation of $ \alpha_j $ for $ j \leq 8 $. The spectral characterization
 of $ \alpha$'s in \cite{beta} allowed a finer computation, reliable for $ j \leq 13 $, and that
 suggested $ \alpha_{j+1} - \alpha_j \simeq 1.515 $ (this is so for the exact potential \eqref{eq:defU} with
 more complicated behaviour for potentials satisfying same symmetries but containing higher
 Fourier modes -- see \S \ref{s:flat}).

 Although it is not clear if the model \eqref{eq:Ham} is physically relevant for larger $ \alpha $'s
 (or, for that matter, if more magic angles exist experimentally) establishing \eqref{eq:quant}
 remains a puzzling mathematical problem. It also remains of interest in physics
 \cite{renu},\cite{nana} but the arguments and the proposed replacements to
 $\frac32$ are not clear.

  One of the difficulties in establishing \eqref{eq:quant} is the ``exponential squeezing" of
 bands proved in \cite[Theorem 4]{beta}: for any $ k \in \mathbb C$ and $ \alpha \gg 1 $,
 there exists $ u \in C^\infty ( \mathbb C/\Gamma ) $ such that
\begin{equation}
\label{eq:Dalph}   \| ( D ( \alpha ) + k ) u \|_{L^2(\mathbb C/\Gamma) } \leq \mathcal O(e^{ - c_0 \alpha }) , \ \ \| u \|_{L^2
 (\mathbb C/\Gamma)}  =1 , \end{equation} 
 that is, we have an almost eigenvalue at every $ k $. This (and a much more precise
 statement) follows from the semiclassical version of H\"ormander's bracket condition -
 see Dencker--Sj\"ostrand--Zworski \cite[Theorem 1.2${}'$]{dsz} and references given there.

In view of \eqref{eq:quant} and \eqref{eq:Dalph}  it is interesting to understand the precise behaviour of the exact solutions to
the eigenvalue problem $ ( D ( \alpha ) + k ) u = 0 $ as
$ \alpha $ gets large (within or without the magic set). As recalled in \S \ref{s:flat} that is
equivalent to the study of the kernel of $ D ( \alpha ) $ on $ H^1 ( \mathbb C/\Gamma ) $.
Numerical simulations -- see \cite[Figure 5]{beta} and the animation
 \url{https://math.berkeley.edu/~zworski/magic.mp4} -- suggest
the presence of regions of exponential decay, $ e^{-c_0 \alpha} $, $ c_0 > 0 $, of the
elements of that kernel.
Although there is no classically forbidden region in the standard sense, some regions
are forbidden in an infinitesimal way explained in Theorem \ref{t:2} below. Our main result
is
\begin{theo}
\label{t:1}
The hexagon spanned by the stacking points
 (see Figure {\rm \ref{f:1}}) has an $ \alpha $-independent neighbourhood $ \Omega $ such that, for some constants $ C_0, c_0 > 0$  any solution of
\begin{equation}
\label{eq:ef}   (  D (\alpha ) + k ) u = 0 , \ \ \  u \in H^1 ( \mathbb C/\Gamma; \mathbb C^2 ) , \ \ \
 \| u\|_{ L^2 ( \mathbb C/\Gamma ; \mathbb C^2 )} = 1 , \end{equation}
satisfies
\begin{equation}
\label{eq:expa}     | u ( z ) | \leq  C_0 e^{ - c_0 \alpha  } , \ \ \  z \in \Omega.
\end{equation}
\end{theo}

\medskip
\noindent

\noindent
{\bf Remarks.} 1. Near the interior of the edges of the hexagon, the theorem is based on a semiclassical Theorem \ref{t:2} below and calculations involving the specific potential. At the corners a more direct argument is used in the Appendix, though the strategy and the method are the same.
A much stronger estimate than \eqref{eq:expa}, valid with all derivatives, holds -- see \eqref{eq:expdec}. It should also be stressed that the result is local and we only need \eqref{eq:ef} to be valid in a neighbourhood of the hexagon.

\noindent
2. An alternative approach to the analysis near the interior of the edges of the hexagon, 
or rather to the underlying microlocal result, 
Theorem \ref{t:3} in \S \ref{s:pr2} (see \eqref{eq:fake}), was recently developed by 
Sj\"ostrand \cite{Sj23}. It provides an attractive alternative to the conjugation method
\eqref{eq:P2D} coming from \cite{sam}.

\noindent
3. The situation is more complicated at the center, $ z_0 $,  of the hexagon, where again
we see exponential decay; there the operator is not even of principal type. In the
notation of Theorem \ref{t:2}, $ dq|_{ q^{-1}(0)\cap \pi^{-1} ( z_0 ) } = 0 $. This suggests that
lower order terms in \eqref{eq:defP} below are important. That is confirmed by the numerical study of
a scalar model based on the leading term in \eqref{eq:defP}: the principal terms agree
but the absence of the lower order term produces no exponential decay at the center -- see
\cite{gaz4}.

The crucial classical (or symplectic geometry) object in the formulation of Theorem \ref{t:2} is the {\em Poisson bracket}: for $f,g \in C^{\infty}(\mathbb R^2_x \times \mathbb R^2_{\xi}) $, where we think of
$ x $ as position and $ \xi $ as momentum, the Poisson bracket is defined as
\[ \{ f, g \} = H_f\, g= \sum_{ j=1}^2 \partial_{\xi_j} f\, \partial_{x_j } g - \partial_{\xi_j} g\, \partial_{x_j } f . \]
Its significance comes from its appearance as the classical observable corresponding
to the commutator of quantizations of $ f $ and $ g $ -- see \cite[(4.3.11)]{zw}.

The manifold  $ \mathbb R^2 \times \mathbb R^2 $ (or $ U \times \mathbb R^2 $ for
$ U \subset \mathbb R^2 $ open) is identified with the (more invariant) cotangent bundle
of $ \mathbb R^2 $,  $T^*\mathbb R^2$ (or $T^* U $). We denote by $\pi: T^* U \rightarrow U$ the natural projection, $\pi(x,\xi) = x$.

\begin{theo}
\label{t:2}
Suppose that
\begin{equation}
\label{eq:defP}    P =  Q  \otimes  I_{\mathbb C^2} + h \begin{pmatrix} R_{11}  &
R_{12} \\ R_{21}  & R_{ 22 }  \end{pmatrix} , \ \ \ x \in U \subset \mathbb R^2 ,\end{equation}
is a principally scalar system of {semiclassical} differential operators with real analytic coefficients in $U$,
$ Q = q ( x, h D_{ x} ) $ is classically elliptic of order {\rm 2}, and $ R_{k\ell} = R_{k\ell}  ( x, h D_{x}) $
are of order {\rm 1}, for $1\leq k,\ell \leq 2$. Suppose that for $ x_0 \in U $, we have
\begin{equation}
\label{eq:bracket}        \{q , \bar q \} |_{ q^{-1} ( 0 ) \cap \pi^{-1} ( x_0 ) } = 0 , \ \ \
\{ q, \{ q, \bar q \}\} |_{ q^{-1} ( 0 ) \cap \pi^{-1} ( x_0 ) } \neq 0 , \end{equation}
and that $ H_{\Re q } $ and $ H_{\Im q } $ are linearly independent on
$ q^{-1} ( 0 ) \cap \pi^{-1} ( x_0 )  $. If $ P u = 0 $ in $ U $ and $ \| u \|_{ L^2 ( U ) } \leq \mathcal O(1)$, then
there exists a neighbourhood $ \Omega $ of $ x_0 $ and $ C_0 , c_0   > 0 $ such that for all $0< h \leq 1$ we have,
\begin{equation}
\label{eq:expdec}
      | \partial^\beta u ( x ) | \leq  C_0 ( |\beta | C_0 )^{|\beta|} e^{ - c_0 / h} , \ \ \ x \in \Omega, \ \
      \beta \in \mathbb N^2 .
\end{equation}
\end{theo}

Theorem \ref{t:1} follows from Theorem \ref{t:2} by considering the operator
\[ P := \alpha^{-2} ( D ( - \alpha ) + k ) ( D ( \alpha + k ) , \ \ \alpha = 1/h . \]
The semiclassical principal symbol of $ P $ is given by the determinant of the
symbol of $ \alpha^{-1} D ( \alpha )$:
\[ q ( x , \xi ) = ( 4 \bar \zeta )^2 - U ( z , \bar z ) U ( -z, - \bar z  ) , \ \ \ z = x_1 + i x_2 , \ \
\bar \zeta = \tfrac12 ( \xi_1 + i \xi_2 ), \]
where we now stress the real analyticity of $ U $ by writing it as the restriction to the totally real submanifold $ \mathbb C \simeq  \{ ( z, \bar z ) : z \in \mathbb C \} \subset \mathbb C^2 $ of a function holomorphic in $ \mathbb C^2 $.
For any fixed $ z $, the range of $ q (x, \xi) $  is $ \mathbb C $ as $ \xi $ varies,
that is, there is no classically forbidden region in the standard sense.
Similarly, if we consider Floquet theory  (see \S \ref{s:flat}) and
look at the eigenfunctions of $ H_k ( \alpha ) := e^{- i \langle k , z \rangle } H( \alpha )
e^{ i \langle k , z \rangle } $ in $ H^1 ( \mathbb C/\Gamma; \mathbb C^4 ) $, we see
that the range of eigenvalues of the symbol on each fiber
$ T^*_z ( \mathbb C /\Gamma ) $ is $ \mathbb R $.

The exponential decay near the hexagon is a consequence of the classical condition \eqref{eq:bracket}
which effectively determines a classically forbidden region for the
eigenfunctions away from the vertices of the hexagon.
The behaviour of $ |\{ q, \bar q \}|_{q^{-1}(0)}  |$ is shown in Figure \ref{f:1}
-- it can be considered as a function of $ z $.
As we already mentioned, \cite[Theorem 4]{beta} shows that near points where the bracket does
not vanish we can construct localized pseudo-modes,
$ ( D ( \alpha ) + k ) u = \mathcal O ( e^{ -c_0 \alpha })$, $ \| u \|_{L^2} = 1 $,
and this is indeed where the actual eigenfunctions concentrate (see
the animation link above). For the animation showing
$ \log |u_K ( \alpha ) | $, where $ u_K $ is the protected state
satisfying  $ ( D ( \alpha ) + K ) u_K = 0 $, $ \alpha = 8 e^{ 2 \pi i \theta } $, $ \| u_K \|_{ L^2
( \mathbb C/\Lambda )} = 1 $, and of the corresponding $ |\{ q, \bar q \}|_{q^{-1}(0)}| $, 
see \url{https://math.berkeley.edu/~zworski/bracket_dynamics.mp4}.

Recently, Sj\"ostrand and Vogel \cite{SV} have also investigated semiclassical properties
of operators for which \eqref{eq:bracket} holds and obtained delicate tunneling estimates in
a model case in which separation of variables was possible. An extension of those results
would likely have consequences for the operators we consider.

Theorem \ref{t:2} is a consequence of the microlocal Theorem \ref{t:3} in \S \ref{s:pr2} and of
the classical ellipticity of the operator $ Q $ (see Proposition \ref{p:global}).
Theorem \ref{t:3}  is in turn a semiclassical version of a theorem of Tr\'epreau \cite{tre} whose
proof relied on ideas and methods introduced by Kashiwara and Kawai \cite{kaka}.
Himonas \cite{him} provided proofs of some of the results of Tr\'epreau using Sj\"ostrand's
approach to analytic microlocal theory \cite{sam}, see also \cite{HiS}.
The results in \cite{tre}, \cite{him} were
proved in the more complicated setting of analytic hypoellipticity
but only for scalar operators.
 Here we are interested
in a purely semiclassical statement which is valid for principally scalar systems. We specialize
to dimension two but the statement and the methods of proof remain valid in all dimensions.
We follow some aspects of \cite{him} but depart from that paper by using the full strength of \cite[Theorem 7.9]{sam}
(see also Tr\'epreau \cite{tre}; the idea of using plurisubharmonic minorants is attributed to
Kashiwara). We also avoid real analytic pseudodifferential operators and Egorov's theorem
for them, absorbing the real canonical transformation into an FBI transform.

We conclude this introduction by reviewing organization of the paper and outlining
some aspects of the proof. In \S \ref{s:red} we study the chiral model starting with
a review of the flat band theory in \S \ref{s:flat} -- this adds to the motivational
discussion above.  We then show how Theorem \ref{t:1}
follows from Theorem \ref{t:2}. In particular we find an explicit formula for
$ \{ q , \{ q, \bar q \}\} |_{ q^{-1} ( 0 ) } $ on the open edges of the hexagon for the
potential given by \eqref{eq:defU}. At the corners of the hexagon, the bracket 
$ \{ q,  \{ q,  \{ q,  \{ q, \bar q \} \} \} \}  |_{ q^{-1} ( 0 ) } $ does not
vanish but all shorter iterations of Poisson brackets are
vanishing. The semiclassical analogue of \cite[Th\'eor\`eme 2]{tre} does not
apply as the inequalities between ``Egorov--H\"ormander numbers" are not satisfied.
By more ad hoc methods based on the specific structure of the symbol $ q $ near
the vertices, that case is covered in the Appendix.

\S 3 is devoted to microlocal preliminaries: definition of the semiclassical (analytic)
wave front set of a distribution $ u $, denoted $ \WFh ( u ) $ here, introduction of general FBI transforms, and a review of the invariance
of the definition of the wave front set. The only slightly nonstandard fact is Proposition \ref{p:phi2psi}
which shows how to obtain FBI transform phase functions compatible with real analytic canonical transformations.
In \S \ref{s:prince} we follow  \cite{him} and obtain a real analytic symplectic reduction
of the symbol to an approximation of a model case $ \xi_1 + i ( x_1^2 + \xi_2 ) $. This follows
a long tradition in the subject -- see \cite[\S 21.3]{H3}.  \S \ref{s:eik} provides
a solution of the complex eikonal equation
associated to the approximate model symbol. That involves a rescaling
similar to that in \cite{him}. 
It also provides the analysis of the associated weights -- see \eqref{eq:eik1} below.

The proof of a microlocal version of Theorem \ref{t:2} is given in \S \ref{s:pr2} and relies on
the analysis in \S \ref{s:eik}.
The goal is to show that
\begin{equation}
\label{eq:fake}  \begin{gathered}   q ( \rho ) = 0 , \ \  dq \! \not \, \parallel \! d\bar q,  \ \ \{ q , \bar q \} (\rho ) = 0 ,   \\
 \{ q, \{ q ,  \bar q \} \} ( \rho ) \neq 0, \ \  \rho \notin  {\rm{WF}}_h ( P u )
 \end{gathered} \ \Bigg \} \ \Longrightarrow \
 \rho \notin {\rm{WF}}_h ( u ) . \end{equation}
For that we use the phase function from \S \ref{s:eik}  to construct an
FBI transform  $ T_h $ (incorporating the canonical transformation from \S \ref{s:prince} using
Proposition \ref{p:phi2psi})
such that, for our system $ P$,
\begin{equation}
\label{eq:P2D}    T_h P \equiv h D_{x_1}  T_h . \end{equation}
(The equivalence here is formulated using exponentially weighted spaces of holomorphic functions, see
\eqref{eq:Sjeq}.) This is done for systems such as \eqref{eq:defP}.

The phase of this new FBI transform satisfies the (holomorphic) eikonal equation
\begin{equation}
\label{eq:eik1}  \partial_{x_1} \varphi ( x, y ) = q ( y , - \partial_y \varphi ( x, y ) ) , \ \ \
\Phi ( x ) := \sup_{y \in \neigh_{\mathbb R^2} (y_0) } - \Im \varphi ( x, y ) , \end{equation}
where $ q $ is the principal symbol of $ Q $ in \eqref{eq:defP} and $ \rho= (y_0, \eta_0 ) \in
T^* \mathbb R^2 $ satisfies the condition in \eqref{eq:fake}. We have for all $ \varepsilon > 0 $,
\begin{equation}
\label{eq:weight2}
| T_h u ( x ) | \leq C_\varepsilon \exp ( ( \Phi ( x ) + \varepsilon  ) / h ), \ \ x \in
\neigh_{\mathbb C^2 } ( x_0 ) , \ \  x_0 = \pi ( \kappa_\varphi ( \rho ) ),
\end{equation}
where $ \kappa_\varphi $ is the complex symplectomorphism associated to $ \varphi $ -- see \eqref{eq54}. The key fact is that the wave front set is independent of the choice of $ \varphi $ -- see \cite[Proposition 2.6.4]{HiS} and Proposition \ref{p:FBI} below. Hence to obtain
\eqref{eq:fake}, we need to show that we have an exponential improvement over \eqref{eq:weight2}, that is, that for some $ \delta > 0 $,
\begin{equation}
\label{eq:wideT}   | T_h u ( x ) | \leq C \exp ( ( \Phi ( x ) - \delta ) / h ), \ \ x \in
\neigh_{\mathbb C^2 } ( x_0 ) , \ \  x_0 = \pi ( \kappa_\varphi ( \rho ) ).\end{equation}

Assuming for simplicity that
$ P u = 0 $ and that $ x_0 = 0 $, \eqref{eq:P2D} shows that
$ T_h u ( x ) $ is essentially independent of $ x_1 $. This means that
the weight $ \Phi ( x ) $ in \eqref{eq:weight2} can be replaced by
its minimum over $ x_1 \in \neigh_{\mathbb C } (0 )$, The key idea, attributed to Kashiwara in \cite{tre} (though implemented there
using different technology), is to prove that for some fixed $ \delta > 0 $ we have, 
\begin{equation}
\label{eq:kash}
\text{  $ \Psi (x_2 ) \leq
\min_{ |x_1 | \leq \varepsilon }  \Phi ( x)  $ and $ \Psi (x_2 ) $ is subharmonic}
\ \Longrightarrow \ \Psi  ( 0 ) \leq \Phi ( 0 ) - \delta .
\end{equation}
That is done in Lemma \ref{l:minorant}.
Applying \eqref{eq:kash} to $ \Psi ( x_2 ) := h \log | T_h u ( 0, x_2 ) | $  gives \eqref{eq:wideT}.

Finally, we pass from the microlocal statement \eqref{eq:fake} (Theorem \ref{t:3})
to an exponential decay statement \eqref{eq:expdec}. That relies on the classical 
ellipticity of the operator $ P$ and is given in Proposition \ref{p:global}. Although
seemingly standard we could not find a reference for the semiclassical case
and relied on recent work by Galkowski--Zworski \cite{gaz1},\cite{gaz2} (based on
\cite{HS}, \cite{Sj96}) to give a  short proof.

The appendix, which is the joint work of Zhongkai Tao and the second-named author, treats the case of the corners of the hexagon, that is, in physical nomenclature, neighbourhoods of stacking points -- see Figure~\ref{f:1}. We follow the same
procedure but use the special structure of $ q = ( 2 \bar \zeta )^2 - U ( z ) U (- z ) $
near $ z = z_S $. That allows an explicit analysis of a solution to \eqref{eq:eik1}  without
taking a preparatory canonical transformation.
We obtain \eqref{eq:kash} for the corresponding weight
and the same method applies.

\medskip

\noindent
{\sc Acknowledgements.} We would like to thank Johannes Sj\"ostrand for
helpful conversations and in particular for directing us to the work of Tr\'epreau \cite{tre} and
to \cite[Theorem 7.9]{sam}.  We are also grateful to Mark Embree for help with Figure~\ref{f:1}
and to Simon Becker for the linked movies.
This work has been partially supported by the
Simons Targeted Grant Award No. 896630, ``Moir\'e Materials Magic" .

\section{Reduction to Theorem \ref{t:2}}
\label{s:red}

We first review some basic facts about symmetries of $ D ( \alpha ) $
and flat bands for the model \eqref{eq:Ham}. We then discuss the reduction to an
operator of the type appearing in Theorem \ref{t:2}. Finally, we show that
bracket conditions \eqref{eq:bracket} hold at the interior points of the edges
of the hexagon spanned by the stacking points (see Figure \ref{f:1}) and discuss
properties of the Poisson brackets at the corners.

\subsection{Flat bands and protected states}
\label{s:flat}
The potential $ U $ in \eqref{eq:defU} enjoys the following symmetries
with respect to the lattice in \eqref{eq:defG}, the rotation by $ 2 \pi/3 $, and
complex conjugation:
\begin{equation}
\label{eq:symU}
U(z  + \gamma ) = e^{ i \langle \gamma, K \rangle } U  (z ), \
\gamma \in \Lambda ,  \quad U  (\omega z ) = \omega U  (z ), \ \
 \overline{U( \bar z ) } = - U ( - z ). \end{equation}
The operator $ D( \alpha ) $ (and the self-adjoint Hamiltonian $ H ( \alpha ) $) are
periodic with respect to $ \Gamma = 3 \Lambda $ and assumptions \eqref{eq:symU}
are enough to guarantee that there exists a discrete set $ \mathcal A \subset \mathbb C $ such
that for $ D ( \alpha ) $ with the domain given by $ H^1 ( \mathbb C/\Gamma ) $,
\[  \Spec_{ L^2 ( \mathbb C/\Gamma ) } D ( \alpha ) = \left\{ \begin{array}{ll}
\Gamma^*, & \alpha \notin \mathcal A , \\
\mathbb C, & \alpha \in \mathcal A , \end{array} \right.  \]
see \cite[Theorem 2]{beta} and for a finer version using the lattice $ \Lambda $,
\cite[Proposition 2.2]{bhz2}. For $ \alpha \notin \mathcal A $,
\[ \dim \ker_{ H^1 ( \mathbb C / \Gamma ) } D ( \alpha ) = 2 , \]
and the spectrum of $ D ( \alpha ) $ is periodic with respect to $ \Gamma^* $ --
see \cite[Theorem 1]{beta} or \cite[Proposition 2.1]{bhz2}.

The bands of the Hamiltonian $ H ( \alpha ) $ are defined as the eigenvalues
of $ H_k ( \alpha) := e^{ - i \langle z, k \rangle } H(\alpha)  e^{ i \langle z , k \rangle } $,
$ \langle z , k \rangle := \Re (\bar z k) $, with the domain $ H^1 ( \mathbb C/\Gamma ) $
(for a finer description see \cite{bhz2}). Since these eigenvalues are symmetric
with respect to $ 0 $ and at $ k = 0 $, the protected states give a multiplicity four eigenvalue
at $ 0 $, we can write the spectrum as
\[ \Spec_{ L^2 ( \mathbb C/ \Gamma ; \mathbb C^4 ) } H_k ( \alpha ) = \{ E_j ( \alpha , k ) \}_{ j \in
\mathbb Z \setminus \{ 0 \} } , \ \ \ E_{\pm j } ( \alpha , 0 ) = 0, \ \ j = 1, 2. \]
Bloch--Floquet theory shows that if we consider $ H ( \alpha ) $ with domain
given by $ H^1 ( \mathbb C; \mathbb C^4 ) $, then
\[  \Spec_{ L^2 ( \mathbb C )} H ( \alpha ) =
\bigcup_{ k \in \mathbb C/\Gamma^* } \{ E_j ( \alpha, k ) \}_{ j\in \mathbb Z \setminus \{ 0 \} } . \]
A flat band at $ 0 $ corresponds to
\begin{equation}
\label{eq:flat}    E_{\pm j } ( \alpha , k ) = 0 \ \text{ for all $ k \in \mathbb C $,  $j = 1,2 $. }
\end{equation}
The definition of $ H_k ( \alpha ) $ above shows that this is equivalent to
$ \Spec_{ L^2 ( \mathbb C/\Gamma ) } D ( \alpha ) = \mathbb C $ and the eigenfunctions are given by
\[ \begin{gathered}     \begin{pmatrix} u (k) \\ u^* ( k ) \end{pmatrix} \in
C^\infty ( \mathbb C / \Gamma ; \mathbb C^4 ) , \\
( D ( \alpha ) + k ) u ( k ) = 0 , \ \ \  ( D ( \alpha )^* + \bar  k ) u^* ( k ) =  0 , \ \  u(k), u^* ( k) \in
C^\infty ( \mathbb C / \Gamma ; \mathbb C^2 ) .
\end{gathered}
\]
The functions $ u ( k ) $ and $ u^* ( k ) $ are easily related to each other (see for instance
\cite[(2.10)]{bhz2}) and in addition the functions $ z \mapsto | u ( k , z) | $ are periodic with respect to the
the small lattice $ \Lambda $ (see \eqref{eq:defG}). Hence when looking for ``classically forbidden"
regions for $ u ( k ) $ (as $ \alpha \to \infty $) we can consider the fundamental domain of $ \Lambda $ shown in  Figure \ref{f:1}. Using the ``theta function argument" (see \cite[\S 3]{bhz2}) $ u ( k ) $ can
be obtained from the $ u ( 0 ) $. Hence, we are effectively looking for classically forbidden regions of the
protected elements of the kernel of $ D ( \alpha ) $ -- see
\url{https://math.berkeley.edu/~zworski/magic.mp4}.

\subsection{Reduction to the principally scalar case}
\label{s:redp}
We start by noting that $ D ( - \alpha ) $ is the (formal) adjugate matrix of $ D ( \alpha ) $ and
\begin{equation}
\label{eq:adj}
\begin{gathered}
( D ( - \alpha ) + k ) ( D ( \alpha ) + k ) = Q_k ( \alpha )  \otimes I_{\mathbb C^2} + \alpha R_k ( \alpha ) =: P_k ( \alpha ),  \\
Q_k ( \alpha ) = (2 D_{\bar z })^2- \alpha^2 U ( z ) U ( - z ) , \\
R_k ( \alpha )  = \begin{pmatrix}   k \alpha^{-1} 4 D_{\bar z } + k^2 \alpha^{-1} &
2 D_{\bar z } U (z ) \\
- ( 2 D_{\bar z } U ) ( -z ) &
 k \alpha^{-1} 4 D_{\bar z } + k^2 \alpha^{-1}\end{pmatrix} .
\end{gathered}
\end{equation}
In particular if $ u \in \ker_{L^2( \mathbb C/\Gamma ) } ( D ( \alpha ) + k ) $ then
$  ( Q_k  ( \alpha ) \otimes I_{\mathbb C^2} + \alpha R_k ( \alpha ) ) u = 0 $.

\noindent
{\bf Remark.} We recall from \cite{beta} that $ \alpha $ is magical if and only if
$ \ker_{L^2( \mathbb C/\Gamma  ) } ( D ( \alpha ) + k ) \neq \{ 0 \} $ for some
$ k \notin \Gamma^* = \frac13  \Lambda^* $. 
 That is equivalent to $ \ker_{L^2 ( \mathbb C/\Gamma ) } P_k ( \alpha ) \neq \{ 0 \} $,
for some $ k \notin \frac13 \Lambda^* $.
In fact, if $ u \in \ker_{L^2( \mathbb C/\Gamma ) } ( D ( \alpha ) + k )  $ then obviously
$ P_k ( \alpha ) u = 0 $. On the other hand if $ P_k ( \alpha ) u = 0 $ and $ v := ( D ( \alpha ) + k ) ) u \neq 0 $ then $ ( D ( - \alpha ) + k ) ) v = 0$. Since
\[   D ( - \alpha ) = - \mathscr R D ( \alpha ) \mathscr R, \ \ \ \mathscr R \begin{pmatrix}
u_1 \\ u_2 \end{pmatrix} ( z ) = \begin{pmatrix} u_2 ( -z ) \\
u_1 ( - z ) \end{pmatrix} , \]
we have $ \mathscr R v \in \ker_{L^2( \mathbb C/ \Gamma ) } ( D ( \alpha ) - k )$, $ -k \notin \frac 13 \Lambda^* $,
that is, $ \alpha $ is magical. \qed

We now take a semiclassical point of view~\cite{zw}, put $ h := 1/\alpha $, and define
\begin{equation}
\label{eq:opP}  \begin{gathered} P = h^2 P_k(\alpha) = Q  \otimes I_{\mathbb C^2}  + h R, \\
Q := ( 2 h D_{\bar z} )^2 - U ( - z ) U ( z ) = q ( x , h D_x ) , \\
q ( x, \xi ) = ( \xi_1 + i \xi_2 )^2 - V ( x ) , \ \ \  V ( x ) =  U (z ) U(-z), \ \ \ z = x_1 + i x_2 ,
\\  R =  \begin{pmatrix}
\ \ \ k 4 hD_{\bar z } + h k^2  & 2D_{\bar z } U ( z )  \\
- (2D_{\bar z } U) ( -z )   & \ \ k 4 hD_{\bar z } + h k^2 \end{pmatrix} , \end{gathered}
\end{equation}
which is the form of the operator in Theorem \ref{t:2}, provided that $k\in \mathbb C$ is confined to a fixed bounded set.
We now verify that assumption \eqref{eq:bracket} is satisfied for $ x_0 $ on the open edges of the hexagon in Figure \ref{f:1}.

\subsection{Bracket computations}

It is convenient to use the complex notation $ \zeta = \frac 12 ( \xi_1 - i \xi_2 ) $,
$ \partial_\zeta = \partial_{\xi_1 } + i \partial_{\xi_2 } $,
$ z = x_1 + i x_2 $, so the real symplectic form on $ T^* \mathbb R^2 $ is given by
\begin{equation}
\label{eq:stand}  2 \Re d \zeta \wedge d z = d \zeta \wedge d z + d \bar \zeta \wedge d \bar z .
\end{equation}
Consequently,
the Poisson bracket is given by
\begin{equation}
\label{eq:defPo} \{ a , b \} = \sum_{ j=1}^2 \partial_{\xi_j }a\, \partial_{x_j} b -
\partial_{\xi_j } b\, \partial_{x_j} a =
\partial_\zeta a\, \partial_z b  - \partial_\zeta b\, \partial_z a + \partial_{\bar \zeta } a\,
\partial_{\bar z } b - \partial_{\bar \zeta } b\, \partial_{
\bar z } a . \end{equation}
We then have (strictly speaking we should write $V = V ( z, \bar z ) $),
\begin{equation}
\label{eq:defq}   q = 4 \bar \zeta^2 - V ( z), \ \  \ V ( z ) =  U ( z ) U ( -z ) , \ \ \
U ( z ) = \lambda i \sum_{ \ell = 0 }^2 \omega^\ell e^{ i \langle z , \omega^\ell K \rangle },
\end{equation}
where $ K = \tfrac43 \pi $, and we introduced a general coupling constant $\lambda$. If $ \lambda \in \mathbb R $ (which is the case for  \eqref{eq:defU}) then,
\begin{equation}
\label{eq:real}
\overline{ U ( \bar z ) } = -  U ( - z ) , \ \ \
\overline{ V (\bar z ) } =V ( z ) = V ( - z ).
\end{equation}
In particular, when $ t \in  \mathbb R $, then $ \Im V ( it ) = 0 $,
$ \Re U ( it ) = 0 $. Also, $ \partial_z U ( z ) = - \frac12 K \lambda \sum_{\ell = 0 }^2 e^{ i \langle z , \omega^\ell K \rangle } $, so
\begin{equation}
\label{eq:reald}
\begin{gathered}
\overline{ (\partial_z U ) ( \bar z  )} = (\partial_z U ) ( - z ) , 
\ \ \ \overline{ \partial_z V ( \bar z ) }  = - ( \partial_z V ) ( - z ) ,
\end{gathered}
\end{equation}
and in particular, $ \Re  \partial_z V ( it ) = 0$ for $ t \in \mathbb R $.

\begin{center}
\begin{figure}
\includegraphics[width=16cm]{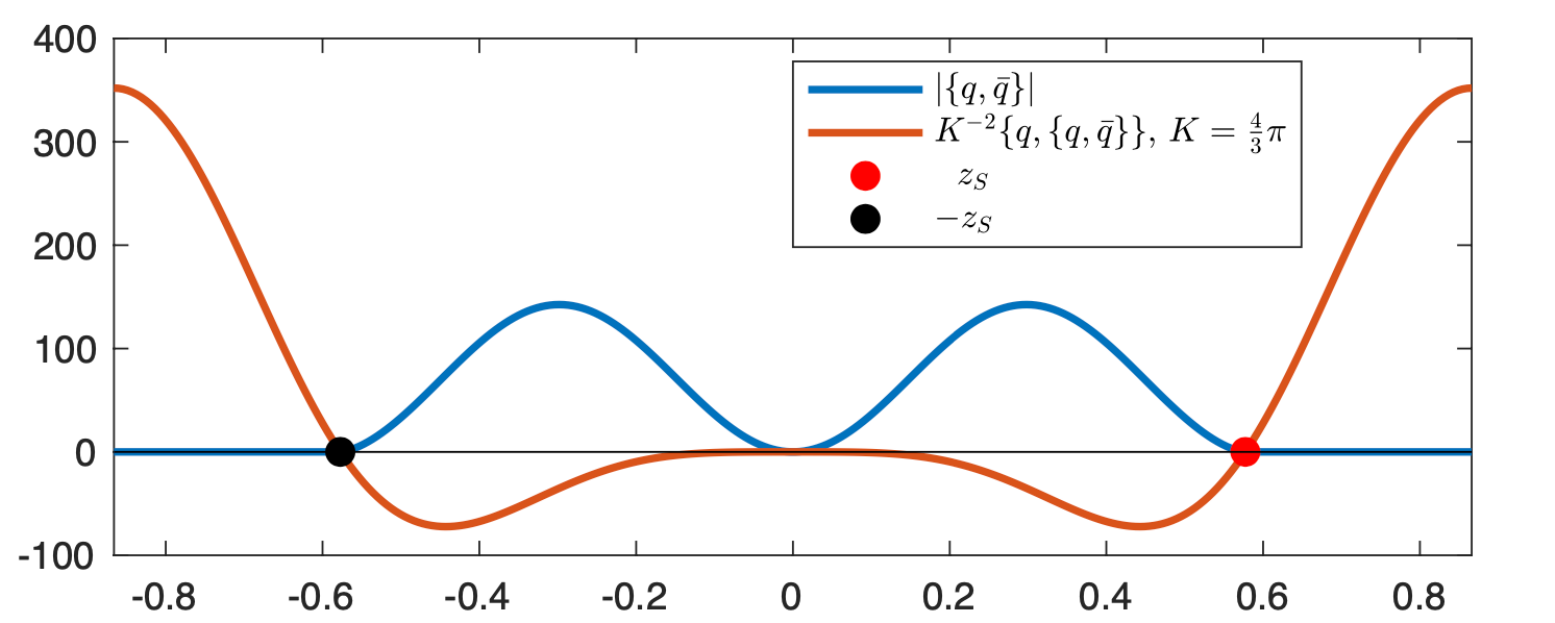}
\caption{Plots of $ |\{ q, \bar q\} | $ and of (rescaled) $ \{ q, \{ q , \bar q \}\} $ above the
intersection of the imaginary axis and the fundamental domain in Figure~\ref{f:1}. The edges of
the hexagon emanate right of $ z_S $ and left of $ - z_S$.
\label{f:bra}}
\end{figure}
\end{center}

The next two lemmas show that the conditions in Theorem \ref{t:2} are satisfied for
the principal symbol of the operator $ P $ in \eqref{eq:opP}. They are illustrated numerically
in Figure~\ref{f:bra}.
\begin{lemm}
\label{l:1}
Let
\[  H := \cup_{\pm } \cup_{ k=0}^2 \pm (1 + \omega^k [0 , \tfrac12] )i/\sqrt 3 + \Lambda , \]
be the hexagon spanned by the stacking points $ \pm z_S + \Lambda $, $ z_S = i/\sqrt 3$
(see Figure {\rm \ref{f:1}}).  Then $ q $ given in \eqref{eq:defq} with $ \lambda \in \mathbb R $ satisfies
\begin{equation}
\label{eq:bracket1}  \{ q , \bar q \} ( \rho ) = 0 \  \text{ if } \  z \in H , \  \rho \in  \pi^{-1} ( z ) \cap q^{-1} ( 0 ) ,
\end{equation}
where $\pi: T^*(\mathbb C/\Lambda) \rightarrow \mathbb C/\Lambda$ is the natural projection, $\pi(z,\zeta) = z$.
\end{lemm}
\begin{proof}
We note some basic symmetries relevant to the hexagon. We define the following
symplectomorphism (for the standard real symplectic form \eqref{eq:stand}; it also preserves
the complex one $ d \zeta \wedge d z $):
\begin{equation}
\label{eq:symR}
R : ( z_S + z , \bar{z}_S + \bar z  , \zeta , \bar \zeta ) \longmapsto ( z_S + \omega z ,
\bar z_S + \bar \omega \bar z , \bar \omega \zeta , \omega \bar \zeta ) .
\end{equation}
Then,
\begin{equation}
\label{eq:symR1}   R^* q = \omega^2 q , \ \ \  R^* \{ q , \bar q \} = \{ q , \bar q \}, \ \ \
R^* \{ q , \{ q , \bar q \} \} = \omega^2 \{ q , \{ q , \bar q \} \} .
\end{equation}
Hence it is enough to check \eqref{eq:bracket1} for $ z \in z_S  [ 1 , \frac32 ]  $. Since
\begin{equation}
\label{eq:Hq}  H_q = 8 \bar \zeta \partial_{\bar z } + \partial_z V \partial_\zeta + \partial_{\bar z } V \partial_{\bar \zeta } ,
\end{equation}  we have, when $ q = 0 $,
\begin{equation}
\label{eq:q2} \{ q , \bar q \} = 8 \zeta \partial_z V  -  8 \bar \zeta \overline{ \partial_{ z }  V} =  16 i \Im
( \zeta \partial_z V) = \pm  8 i \Im \left( ( \overline{V ( z ) })^{\frac12}  \partial_z V ( z ) \right)  ,
\end{equation}
where $ \pm $ depends on $ 2\bar \zeta = \mp ( V ( z ) )^{\frac12} $ (so that $ q = 0 $).
From \eqref{eq:real} and  \eqref{eq:reald} we see that $ V $ is real and $ \partial_z V $ is purely imaginary
on $ i \mathbb R $,
and it remains to show that $ V ( i t) < 0 $ for
$  1/\sqrt 3 < \pm t \leq \sqrt 3/2 $. For that we calculate
\[   U ( i t ) = i \lambda \sum_{ \ell = 0}^2 \omega^\ell e^{ \frac i 2 K ( it \bar \omega^\ell - i t \omega^\ell ) }
=
\lambda i ( 1 + 2 \cos( 2 \pi ( t \sqrt 3 + 1 )/3 ) ).  \]
This means that we need
\begin{equation}
\label{eq:coscos}
\left( 1 + 2 \cos\left( {2 \pi} ( 1 + t \sqrt 3)/3 \right)\right) \left( 1 + 2 \cos\left( 2 \pi( 1 - t \sqrt 3)/3 \right)\right) > 0,
\end{equation}
for $ 1/\sqrt 3 <  t \leq \sqrt 3/2$. Writing
$
2 \pi ( 1 + t \sqrt 3) /3 = \pi + \theta$, $ {\pi}/{3} < \theta \leq 2\pi/{3}$,
it follows that
$$
1 + 2 \cos\left( {2 \pi} ( 1 + t \sqrt 3)/3 \right) = 1 -2 \cos \theta > 0.
$$
Similarly, with
$
{2 \pi} ( 1 - t \sqrt 3)/3  = \theta$ , $ -{\pi}/{3} \leq \theta < 0,
$
we obtain
$$
1 + 2 \cos\left( {2 \pi} ( 1 - t \sqrt 3)/3 \right) = 1 + 2\cos \theta > 0,
$$
and \eqref{eq:coscos} follows.
\end{proof}

\begin{lemm}
\label{l:2}
For the potential given in \eqref{eq:defU} and for $ z \in \pm( z_S + \omega^k (0, \frac12 z_S ] ) $,
$ z_S = i / \sqrt 3 $, $0\leq k \leq 2$ (open edges of the hexagon),  $H_{{\rm Re}\,q}(\rho)$, $H_{{\rm Im}\, q}(\rho)$ are linearly independent for
$\rho \in \pi^{-1} ( z ) \cap q^{-1} ( 0)$. Furthermore,
\begin{equation}
\label{eq:tripleb}  \omega^{2k} \{ q , \{ q, \bar q \} \} ( \rho ) > 0 , \ \ \ \rho \in \pi^{-1} ( z ) \cap q^{-1} ( 0).
\end{equation}
\end{lemm}
\begin{proof}
As in the proof of Lemma \ref{l:1}, we use \eqref{eq:symR1} so that it is enough to consider $ z = it $, $
1/\sqrt 3 < \pm t \leq \sqrt 3/2 $. First, the linear independence of the Hamilton vector fields $H_{{\rm Re}\,q}(\rho)$, $H_{{\rm Im}\, q}(\rho)$ follows from (\ref{eq:Hq}), combined with the fact that $ V ( i t ) < 0 $ for $ 1/\sqrt 3 < \pm t \leq \sqrt 3/2 $, established in the proof of Lemma \ref{l:1}. Next, from \eqref{eq:Hq} we see that, at points $\in q^{-1}(0)\cap \pi^{-1}(it)$,
\begin{equation}
\label{eq:3q}  \begin{split}  \{ q , \{ q , \bar q \} \} & =
( 8 \bar \zeta \partial_{\bar z } + \partial_z V \partial_\zeta + \partial_{\bar z } V \partial_{\bar
\zeta } ) ( 8 \zeta \partial_z V  -
8 \bar \zeta \overline{ \partial_{ z }  V}  )
 \\ & = 64 (
  |\zeta|^2  \partial_z \partial_{\bar z}  V - \bar \zeta ^2 \overline{ \partial_z^2 V } ) +  8 ( (\partial_z V) ^2 - \partial_{\bar z} V \overline { \partial_z V } ).
\\
& = 16 (    | V | \partial_{z} \partial_{\bar z}  V -  V  \overline {\partial_z^2 V }) + 8 (( \partial_z V )^2 -  \partial_{\bar z } V \overline { \partial_z V } ) \\
& = - 16 V (  \partial_{z} \partial_{\bar z}  V + \overline {\partial_z^2 V } ) +
8  (( \partial_z V )^2 -  \partial_{\bar z } V \overline { \partial_z V } ) .
\end{split}
\end{equation}
where we again used that $ V ( i t ) < 0 $ for $ 1/\sqrt 3 < \pm t \leq \sqrt 3/2 $. A computation based on \eqref{eq:defq} shows that
\begin{equation}
\label{eq:ddU} \begin{gathered}  \partial_{\bar z } U ( z )
= \tfrac 12 i K U ( \bar z ) , \ \ \
 \partial_z^2 U ( z )
=  - \tfrac14 K^2 U ( \bar z ) .
\end{gathered} \end{equation}
Hence, using also
$ \partial_z \left(U(\bar{z})\right) = (\partial_{\bar{z}} U)(\bar{z})$ and $  \partial_z \left(U(-\bar{z})\right) = -(\partial_{\bar{z}} U)(-\bar{z})$, we obtain
\[ \begin{split} \partial_z V ( z ) &= \partial_z U ( z ) U ( -z ) - (\partial_z U ) ( - z ) U ( z )  , \\
\partial_{\bar z } V ( z ) & =  \partial_{\bar z }  U ( z ) U ( -z ) - (\partial_{\bar z } U ) ( - z ) U ( z )
=  \tfrac12  i K \left( U ( \bar z ) U ( - z ) - U ( - \bar z ) U ( z )  \right) , \\
\partial_{z \bar z  }^2 V ( z ) & =
 \tfrac12 iK \left( (\partial_{\bar z } U ) ( \bar z ) U ( -z ) - U ( \bar z ) (\partial_{z } U ) ( - z ) +
( \partial_{ \bar z } U ) ( - \bar z ) U ( z ) - U ( - \bar z ) \partial_z U ( z ) \right) \\
& = -\tfrac12 K^2 U ( z ) U ( -z ) - \tfrac12 i K \left( (\partial_z U ) (- z) U ( \bar z ) + \partial_z U ( z )
U ( - \bar z ) \right), \\
\partial_z^2 V ( z ) & = \partial_z^2 U ( z ) U ( - z ) + ( \partial_z^2 U ) ( -z )  U ( z ) - 2
\partial_z U ( z ) ( \partial_z U ) ( - z ) \\
& = -\tfrac14 K^2 \left( U ( \bar z ) U ( - z ) + U ( - \bar z ) U ( z ) \right)  - 2
\partial_z U ( z ) ( \partial_z U ) ( - z ) .
\end{split} \]
We now put
\begin{equation}
\label{eq:deffg}
\begin{gathered}    f ( t ) := U ( it ) , \ \   g ( t ) := \partial_z U ( it ) , \ \ \ \  \bar f ( t ) = - f ( t ) , \ \  \bar g ( t ) = g ( t ) , \ \ g ( t ) = g ( - t)  \\
f ( t ) = i \lambda ( 1 + 2 \cos ( 2 \pi ( t \sqrt 3 + 1)/3 ) ),  \ \ \
g ( t ) = - \tfrac12 K \lambda ( 1 + 2 \cos ( 2 \pi t \sqrt 3 / 3 )) .
\end{gathered}
\end{equation}
To simplify notation further we take (as we may) $ \lambda = 1 $ and denote
$ c:= \cos ( 2 \pi  t \sqrt 3 /3 ) $ and $ s := \sqrt 3 \sin ( 2 \pi  t \sqrt 3 /3 ) $, so that
\[ g ( t ) = - \tfrac12 K ( 1 + 2c ) , \ \  f( t) = i ( 1 - c - s ) , \ \ f ( -t ) = i ( 1 - c + s ) . \]
Then, using the fact that $ s^2 = 3 ( 1 - c^2 ) $,
a lengthy computation gives
\[  \begin{split} \tfrac 18  K^{-2}  \{ q , \{ q , \bar q \} \}
&  =  ( c- 1  )^2 (  2c + 1 ) ( 2c - 9 )  .
\end{split} \]
For $ 1/\sqrt 3 < t \leq \sqrt 3/2 $, $  - 1 \leq c =\cos ( 2 \pi \sqrt 3 t/3 ) < -\frac12 $. Hence
$ c - 1 , 2c + 1 ,  2c - 9 < 0 ,   $ and $ \{ q , \{ q, \bar q \} \} > 0 $, as claimed.
\end{proof}

We conclude this section by recording the behaviour of the Poisson brackets at the vertices of the hexagon, where Theorem \ref{t:2} does not apply. We note
first that
\begin{equation}
\label{eq:UzS}
\begin{gathered}
U ( z_S ) = 0 , \ \ \partial_z U ( z_S ) = 0,   \ \  \partial_{\bar z } U ( z_S )
= \tfrac{8} 3 \pi^2 ,
\\
U ( - z_S ) = - 4 \pi i , \ \  \partial_z U ( - z_S ) = 0 , \ \  \partial_{\bar z } U ( - z_S)
= 0 .
\end{gathered}
\end{equation}
Hence,
\[  \pi^{-1} ( \pm z_S ) \cap q^{-1} ( 0 ) = \{ ( \pm z_S , 0 ) \}  ,  \ \ \
 dq ( \pm z_S , 0 ) = \pm \tfrac {32} 3 \pi^3 i \, d \bar z \neq 0  . \]
\begin{lemm}
\label{l:vert}
We have
\begin{equation}
\label{eq:Hq3}  H_{q_{j_1}} \cdots H_{q_{j_p} }   \bar q\, ( \pm z_S, 0 ) = 0 , \  p < 4 , \ \
\forall \, j_\ell \in \{ 1, 2 \} , \ \  q_1 = q, \ q_2 = \bar q,   \end{equation}
and
\begin{equation}
\label{eq:Hq4}  H_q^4 \bar q\, (\pm z_S,0) =   \{ q , \{ q , \{ q , \{ q , \bar q \}\}\}\} ( \pm z_S , 0 ) \neq 0 .
\end{equation}
\end{lemm}
\begin{proof}
Since $ q $ is even in $ z $ it is enough to consider $ z_S $. Using \eqref{eq:ddU} and \eqref{eq:UzS}
we see that
that  $ V ( z + z_S ) = U ( -z_S ) \partial_{\bar z } U( z_S ) \bar z +  \frac{ 1}{2}  U ( - z_S )\partial_z^2 U ( z_S ) z^2 + \mathcal O ( |z|^3 ) $, so that
\begin{equation}
\label{eq:Tayq}   p ( z, \zeta ):= \tfrac18  q ( z_S + z , \zeta ) = \tfrac12 \bar \zeta^2  - a \bar z - \tfrac12 b z^2 + \mathcal O ( |z|^3 ) , \ \ \
a =  -\tfrac{4}3 \pi^3 i , \ \ b = \tfrac{8}{9} \pi^4 .
\end{equation}
Consequently,
\[ H_p = \bar \zeta \partial_{\bar z } + a \partial_{\bar \zeta} +  b z \partial_{\zeta} +
\mathcal O ( |z|^2 ) \partial_{\zeta} +\mathcal O ( |z|^2 ) \partial_{\bar \zeta} . \]
Then  $ H_p \bar p = - b \bar z \bar \zeta +  b z \zeta + \mathcal O ( | z |^2 | \zeta |  ) $, and
\begin{gather*}
\begin{split}
H_p^2 \bar p & =
- b \bar \zeta ^2 -  ab \bar z
+ \mathcal O ( |z| |\zeta|^2 + |z|^2 ) , \\
H_p^3 \bar p & = 
-  3 ab \bar \zeta + \mathcal O ( |z ||\zeta| +  |\zeta|^3 + | z |^2) ,
\\
H_p^4 \bar p & = 
- 3 a^2 b + \mathcal O ( |\zeta| + |z| ) \neq 0 , \ \  0 \leq |z|, |\zeta| \ll 1.
\end{split}
\end{gather*}
This proves \eqref{eq:Hq4} and part of \eqref{eq:Hq3}. Since $ H_p \bar p $ is purely
imaginary,  we have,
\[ H_{\bar p } H_{p } \bar p  =  - \overline{H_p^2 \bar p} , \ \ \
 H_{\bar p }^2 H_{p} \bar p = - \overline{ H_p^3 \bar p} , \ \ \
 H_p H_{\bar p}  H_p \bar p  = - \overline{ H_{\bar p } H_p^2  \bar p  }
  \]
so the only remaining case to check is $ H_{\bar p} H_p^2 \bar p $, and that is again straightforward.
\end{proof}

\section{Microlocal preliminaries}
\label{s:sym}

We review aspects of microlocal analysis needed in the proof of Theorem 2.

\subsection{Analytic symbols}
In the semiclassical setting we work locally and consider functions defined in open sets
(typically neighbourhoods of fixed points) $ \Omega \subset \mathbb C^N $. For a continuous
function $ \Phi : \Omega \to \mathbb R $ (typically plurisubharmonic), Sj\"ostrand spaces
(following terminology of Lebeau~\cite{leb}), $ H_\Phi(\Omega) $,
 are defined as spaces of functions $u: \Omega \times ( 0, 1 ]
\to \mathbb C$, satisfying
\begin{equation}
\label{eq:Sjsp}   u ( \bullet , h ) \in \mathscr O (\Omega ) , \ \
\forall \, K \Subset \Omega , \ \varepsilon > 0 \ \exists \, C \ \forall \, h \in ( 0,1 ] \ \forall \, z \in K \ \
| u ( z, h ) | \leq C e^{ ( \Phi ( z )+\varepsilon)/h }.
\end{equation}
For $ z_0 \in \Omega$  and $ \Phi \in C ( \Omega ) $ we also define the space of germs at
$ z_0 $:
\begin{equation}
\label{eq:Sjsp0}
H_{\Phi, z_0 } := \bigcup_{ \Omega'= \neigh_{\mathbb C^N }( z_0 ) \subset \Omega } H_\Phi ( \Omega' ) .
\end{equation}
The equivalence relations on these spaces are given as follows:
\begin{equation}
\label{eq:Sjeq}
\begin{gathered}
u \equiv_{H_\Phi ( \Omega ) }  v  \ \Longleftrightarrow \
\exists \,  \Phi_0 \in C ( \Omega ), \ \Phi_0 < \Phi \ \  u - v \in H_{\Phi_0 }(\Omega), \\
u \equiv_{H_{\Phi,z_0 } }  v \ \Longleftrightarrow  \ \exists \, \Omega' = \neigh_{\mathbb C^N } ( z_0 ) \subset \Omega,  \  u, v \in H_\Phi ( \Omega' ) \ \text{and} \
\ u \equiv_{ H_\Phi ( \Omega' ) } v .
\end{gathered}
\end{equation}
When the context is clear we may drop $ \bullet $ in $ \equiv_\bullet $ or write $ \equiv $ in $ \bullet $.

Analytic symbols in $ \Omega $ are defined using $ \Phi = 0 $: they are given by the space
$ H_0( \Omega ) $. A (formal) classical analytic symbol in $ \Omega $ is a (formal) expression
\begin{equation}
\label{eq:classs}
\begin{gathered}
a ( z; h ) := \sum_{ k = 0 }^\infty h^k a_k ( z ) , \ \  a_k \in \mathscr O ( \Omega ) , \\
 \forall \, K \Subset \Omega \,\,
\exists \, C = C ( K ) \  \ | a_k ( z ) | \leq C^{ k+1} k^k , \ z \in K,\,\,k= 0,1,2,\ldots.
\end{gathered}
\end{equation}
For open $ \Omega_1 \Subset \Omega $ we have a {\em realization} of $ a (z; h ) $ given by
the following holomorphic function:
\begin{equation}
\label{eq:reali}
a_{\Omega_1 } ( z; h ) := \sum_{k=0}^{ [( e C( \Omega_1) h )^{-1} ]} a_k ( z ) h^k ,  \ \
 \Omega_1 \subset \Omega_2 \Subset \Omega
\  \Rightarrow \ a_{\Omega_1} \equiv a_{ \Omega_2 } \text{ in } H_{0} ( \Omega_1 ) .
 \end{equation}
We refer to  \cite[\S 1]{sam} or \cite[\S 2.2]{HiS} for a detailed account.
We recall however the following fundamental result of Boutet de Monvel--Kr\'ee. The composition
of symbols defined there is motivated by the composition formula for pseudodifferential
operators.

\begin{prop}
\label{p:BdMK}
For $ a = a ( x, \xi; h) $ and $ b = b ( x, \xi; h) $,  formal classical analytic symbols in $ \Omega
\subset \mathbb C_x^n \times \mathbb C_\xi^n $ we define
\begin{equation}
\label{eq:prod}  a \# b ( x, \xi; h ) := \sum_{\alpha \in \mathbb N^n } \frac{1} { \alpha!} \partial_\xi^{\alpha} a ( x, \xi; h )
( h D_x )^\alpha b ( x, \xi; h ) . \end{equation}
Then $ a \# b $ is a classical analytic symbol in $ \Omega $. If
$ a \neq 0 $ on $ \Omega $ and $ \Omega_0 \Subset \Omega $ is an open set, then
the formal classical symbol $ b $ defined by
\begin{equation}
\label{eq:BdMK}
a \# b = 1
\end{equation}
is a formal classical analytic symbol in $ \Omega_0 $.
\end{prop}
The condition that $ a \neq 0 $ in $ \Omega $ is referred to as the {\em ellipticity} of $ a $ in
$ \Omega $.

\subsection{Analytic semiclassical wave front sets}
\label{s:WF}

{Let $U$ be an open set in $ \mathbb R^n$}, and suppose that ${(0,h_0]\ni} h \mapsto u ( h ) \in \mathscr D' ( U; \mathbb C^p ) $
is a family of vector-valued $h$-tempered distributions in the sense that for every $ \omega \Subset U $ there exists
$ N $ such that $ \| u ( h ) \|_{H^{-N} ( \omega ) } = \mathcal O ( h^{-N}) $. We follow \cite[Definition 3.2.1]{mart} and define the semiclassical analytic wave front set, $ \WFh (u) \subset T^*U$, as follows:
\begin{equation}
\label{eq:WFa}
\begin{gathered}
( y_0 , \eta_0) \notin \WFh ( u ) \ \Longleftrightarrow \ \left\{ \begin{array}{l}  \exists \, \delta > 0 , C > 0 ,   \ \Omega = \neigh_{ \mathbb C^n } ( y_0 - i \eta_0 ) \\
| T_h u  ( x ) | \leq C e^{ ( \Phi_0 ( x ) - \delta ) /h }  , \ x \in \Omega,\,\, 0 < h \leq h_0,
\end{array} \right.
 \\
T_h w ( x ) :=  \int_{ \mathbb R^n } w ( y ) \chi( y )  e^{ i \varphi_0 ( x, y ) /h } dy , \ \ \varphi_0 ( x, y ) :=
i ( x - y)^2/2 , \ \
\Phi_0 ( x ) := \tfrac12 | \Im x|^2,
\end{gathered}
\end{equation}
where $ \chi \in C^\infty_{\rm{c}} ( U ) $, $ \chi ( {y}) \equiv 1 $ in a neighbourhood of $ y_0 $.
(We should note that for $ h$-independent distributions $ u $ this gives the {\em analytic}
wave front set $ \WF_{\rm{a}} ( u ) $, \cite[\S 8.4]{H1}; for the $ C^\infty $ version in the semiclassical setting
see \cite[\S 8.4]{zw}.)

\subsection{FBI transforms}
\label{s:FBI}

We follow \cite{sam}, \cite[Chapter II]{HiS} to define generalized FBI transforms and prove an essentially well known result about the composition of complex canonical transformations associated to FBI transforms with real canonical transformations.

Generalized FBI transforms are defined using phase functions generalizing
$$
\varphi_0 ( x, y ) := i ( x - y )^2/2,
$$
as follows. We assume that
 $\varphi(x,y)$ is  holomorphic in a neighbourhood of $(x_0,y_0)\in \mathbb C^n \times \mathbb R^n$,
 and that
\begin{equation}
\label{eq52}
-\varphi'_y(x_0,y_0) = \eta_0 \in \mathbb R^n, \quad
{\rm det}\, \varphi''_{xy}(x_0,y_0) \neq 0, \quad {\rm Im}\, \varphi''_{yy}(x_0,y_0) > 0.
\end{equation}
This phase function defines a complex canonical transformation:
\begin{equation}
\label{eq54}
\begin{gathered}
\kappa_{\varphi}: {\rm neigh}_{\mathbb C^{2n}} (y_0,\eta_0) \to {\rm neigh}_{\mathbb C^{2n}}
(x_0,\xi_0), \ \
\xi_0 := \varphi'_x(x_0,y_0) , \\
\kappa_{\varphi} :  ( y, - \varphi_y' ( x, y ) ) \mapsto ( x, \varphi_x' ( x, y ) ).
\end{gathered}
\end{equation}
The image of a real neighbourhood of $ ( y_0, \eta_0 ) $,
$ \Lambda_{\Phi} := \kappa_{\varphi}({\rm neigh}_{\mathbb R^{2n}} (y_0,\eta_0) ) $,
is given by
\begin{equation}
\label{eq6}
\Lambda_{\Phi} = \left\{\left(x,\tfrac{2}{i} {\partial_x  \Phi} (x)\right); x\in {\rm neigh}_{\mathbb C^n} (x_0)\right\} \subset T^*\mathbb C^{n}, \ \ \Phi(x) := \underset{y}{{\rm sup}} (-{\rm Im}\,\varphi(x,y)),
\end{equation}
where the supremum is taken over a small real neighbourhood of $y_0$. The real analytic function $\Phi$ is strictly plurisubharmonic in a neighbourhood of $x_0$ and the manifold $\Lambda_{\Phi}$ is I-Lagrangian and R-symplectic. This means that the restriction of the complex symplectic $(2,0)$--form
$ \sigma = \sum_{j=1}^n d\xi_j \wedge dx_j$ 
on $T^*\mathbb C^n$ to $\Lambda_{\Phi}$ is real non-degenerate. Letting
$ \sigma_{{\mathbb  R}} = \sum_{j=1}^n d\eta_j \wedge dy_j $
be the standard symplectic form on $T^*\mathbb R^n_y$ and writing $\sigma_{\Phi} = \sigma|_{\Lambda_{\Phi}}$, we obtain that the map $\kappa_\varphi $ in \eqref{eq54}
 can be viewed as a canonical transformation between real symplectic manifolds,
\begin{equation}
\label{eq9}
\kappa_\varphi : ({\rm neigh}_{T^*\mathbb R^{n}} (y_0,\eta_0),\sigma_{{\mathbb R}}) \rightarrow (\Lambda_{\Phi},\sigma_{\Phi}).
\end{equation}

We recall the key result \cite[Proposition 7.4]{sam}, \cite[Proposition 2.6.4]{HiS} which shows that
the definition \eqref{eq:WFa} of $\WF_h(u)$ is independent of the choice of an FBI transform:
\begin{prop}
\label{p:FBI}
Suppose that $ u ( h ) \in \mathscr D' ( U; \mathbb C^p ) $ is {an} $h$-tempered family of vector-valued distributions in the
sense of \S {\rm \ref{s:WF}} and that for $ ( x_0 , y_0 ) \in \mathbb C^n \times U $, $ \varphi ( x, y ) $ satisfies \eqref{eq52}. Then $(y_0,\eta_0) \notin \WF_h(u)$ if and only if \eqref{eq:WFa} holds with $ \Omega = \neigh_{\mathbb C^n} ( x_0 ) $, $\Phi_0(x)$ replaced by $\Phi(x)$, and $T_h$ given by
\begin{equation}
\label{eq:FBI}  T_h w ( x )  := \int_{\mathbb R^n} \chi ( y ) w ( y ) a ( x, y; h ) e^{ i \varphi ( x, y )/h } dy  ,
\end{equation}
where $a(x,y;h)$ is an elliptic (matrix valued) classical analytic symbol defined in a neighbourhood of $(x_0,y_0)$, and $ \chi \in C^\infty_{\rm{c}} ( U )$
satisfies $ \chi|_{ \neigh_{\mathbb R^n}(y_0 ) } \equiv 1  $.
\end{prop}

\medskip
\noindent
We now consider a real analytic canonical transformation,
\begin{equation}
\label{eq55}
\kappa : {\rm neigh}_{\mathbb R^{2n}} (0,0) \rightarrow {\rm neigh}_{\mathbb R^{2n}}(y_0,\eta_0), \ \ \
\kappa ( 0, 0 ) = ( y_0, \eta_0 ) .
\end{equation}
We will need the following essentially known result --  see~\cite[Section 1]{SjHokk} for a related discussion in the linear case
(that is the case in which $ \varphi ( x, y ) $ is quadratic).

\begin{prop}
\label{p:phi2psi}
There exists a holomorphic function $\psi(x,z)$ near $(x_0,0)\in \mathbb C^n \times\mathbb R^n$, satisfying
\begin{equation}
\label{eq56}
-\psi'_z(x_0,0) = 0, \quad
{\rm det}\, \psi''_{xz}(x_0,0) \neq 0, \quad {\rm Im}\, \psi''_{zz}(x_0,0) > 0,
\end{equation}
such that,  in the notation of \eqref{eq54}, $
\kappa_{\varphi}\circ \kappa = \kappa_\psi$,
\begin{equation}
\label{eq58}
\begin{gathered}
\kappa_{\psi}: {\rm neigh}_{\mathbb C^{2n}} (0,0) \to  {\rm neigh}_{\mathbb C^{2n}}(x_0,\xi_0) , \ \
\kappa_{\psi} : (z,-\psi'_z(x,z)) \mapsto (x,\psi'_x(x,z)) .
\end{gathered}
\end{equation}
\end{prop}
\begin{proof}
The holomorphic canonical transformation
\begin{equation}
\label{eq59}
\kappa_1 := \kappa_{\varphi}\circ \kappa: {\rm neigh}_{ \mathbb C^{2n}}(0,0) \rightarrow {\rm neigh}_{ \mathbb C^{2n}} (x_0,\xi_0),
\end{equation}
is a diffeomorphism from a real neighbourhood of $(0,0) $ to a neighbourhood of $(x_0,\xi_0)$ in $\Lambda_{\Phi}$, $
\kappa_1({\rm neigh}_{\mathbb R^{2n}}(0,0)) = {\rm neigh}_{\Lambda_{\Phi}}(x_0,\xi_0)
$.
Writing $\kappa_1(z,\zeta) = (x(z,\zeta), \xi(z,\zeta))$, we claim first
that
\begin{equation}
\label{eq60}
\det \partial_\zeta  x (0,0)\neq 0.
\end{equation}
When verifying (\ref{eq60}), it suffices to check that the complex linear canonical transformation $d\kappa_1(0,0)$ satisfies
\begin{equation}
\label{eq61}
d\kappa_1(0,0)(0,\zeta) = ( \partial_\zeta x (0,0) \zeta , \partial_\zeta  \xi (0,0) \zeta ) = (0,\xi) \Longrightarrow     \zeta =  0.
\end{equation}
We first note that
\begin{equation*}
d\kappa_1(0,0)(\mathbb R^{2n}) = T_{(x_0,\xi_0)}\Lambda_{\Phi} = \left\{(\delta_x,\delta_{\xi})\in \mathbb C^{2n}; \delta_{\xi} = \tfrac{2}{i}\left(\Phi''_{xx}(x_0)\delta_x + \Phi''_{x\bar{x}}(x_0)\bar{\delta_x}\right)\right\}.
\end{equation*}
Let $ \iota_\Phi = \iota_{\Phi}(x_0): \mathbb C^{2n} \rightarrow \mathbb C^{2n}$ be the unique anti-linear involution which is equal to the identity on the maximally
totally real linear space $T_{(x_0,\xi_0)}\Lambda_{\Phi}\subset \mathbb C^{2n}$ (see \cite[\S 1.2]{HiS} for a review of these concepts).
It is given by, writing $\Phi'' = \Phi''(x_0)$,
\[   \iota_{\Phi}  ( x , \xi ) =
( - (\Phi''_{ \bar x x })^{-1}  \Phi'' _{\bar x \bar x } \bar x - \tfrac i 2 (\Phi''_{ \bar x x })^{-1}  \bar \xi ,
( \tfrac  2 i \Phi''_{ x \bar x } - {\tfrac 2 i} \Phi''_{xx} (\Phi''_{\bar x  x})^{-1}   \Phi''_{\bar x \bar x}  ) \bar x -  \Phi''_{xx}
(\Phi''_{\bar x x})^{-1}  \bar \xi ).\]
The strict plurisubharmonicity of $ \Phi $ (that is, the strict positivity of $ \Phi''_{x \bar x } $) shows that
\begin{equation}
\label{eq63}
\begin{split}
\tfrac{1}{i}\sigma((0,\xi), \iota_{\Phi}(0,\xi)) & =
\tfrac{1}{i}\sigma( (0, \xi ) , ( - \tfrac i 2 (\Phi''_{ \bar x x })^{-1}  \bar \xi ,
-\Phi''_{xx} (\Phi''_{\bar x x})^{-1} \bar \xi))
\\ &  =  - \tfrac12
\langle  (\Phi''_{  x \bar x })^{-1} \xi, \xi \rangle  < 0 ,\quad 0\neq \xi \in \mathbb C^n,
\end{split}
\end{equation}
(Here, and elsewhere, if $ w , z \in \mathbb C^n $, $ \langle w, z \rangle := \sum_{j=1}^n
w_j \bar z_j $.)
 We also have  $d\kappa_1(0,0) \circ \Gamma = \iota_{\Phi}(x_0) \circ d\kappa_1(0,0)$, where
 $\Gamma: \mathbb C^{2n} \rightarrow \mathbb C^{2n}$ is the complex conjugation map. Combining (\ref{eq63}) with the fact that $
\frac{1}{i}\sigma((0,\zeta), \Gamma(0,\zeta)) = \frac{1}{i}\sigma((0,\zeta), (0, \bar \zeta))= 0$, $ \zeta \in \mathbb C^n,
$
we conclude that (\ref{eq61}) holds, and therefore we obtain (\ref{eq60}).
From that,  the holomorphic implicit function theorem, and the fact that $ \kappa_1 $ is a canonical transformation
 we obtain the existence of a holomorphic function $\psi(x,z)$ in a neighbourhood of $(x_0,0)$ such that $\kappa_1 = \kappa_{\psi}$ in (\ref{eq58}).

The first two conditions in (\ref{eq56}) hold since $\kappa_\psi $ is a canonical transformation. It only remains  to check  the third condition in (\ref{eq56}). For that we observe that the differential of $\kappa_{\psi}$ at $(0,0)$ is given by
\begin{equation}
\label{eq63.1}
d\kappa_{\psi}(0,0): (\delta_z, -\psi''_{zx}\delta_x - \psi''_{zz}\delta_z) \mapsto (\delta_x, \psi''_{xx}\delta_x + \psi''_{xz}\delta_z),\quad \psi'' = \psi''(x_0,0),
\end{equation}
We then consider the complex Lagrangian plane
\begin{equation}
\label{eq64}
\begin{gathered}
V = V_{x_0} := \{(\delta_z, -\psi''_{zz}\delta_z);\, \delta_z \in \mathbb C^n\} = d\kappa_{\psi}(0,0)^{-1}(T^*_0\mathbb C^{n}),  \\ T^*_0 \mathbb C^n = \{ ( 0, \xi ) : \xi \in \mathbb C^n \} \subset T_{ ( x_0 , \xi_0)} ( \mathbb C^{2n} )
\end{gathered} \end{equation}
and note that for $ \xi := \psi_{xz }'' \delta_z $, \eqref{eq63} shows that
\begin{equation}
\label{eq:negL}
\begin{split}
\tfrac 1 i \sigma ( (\delta_z, -\psi''_{zz} \delta_z) , \Gamma ( \delta_z , - \psi_{zz}'' \delta_z )   ) &  =
\tfrac 1i \sigma (d\kappa_{\psi}(0,0)^{-1} ( 0 , \xi ) , d\kappa_{\psi}(0,0)^{-1} \iota_\Phi  ( 0 , \xi ) )
\\ &  = \tfrac 1i \sigma ( ( 0 , \xi ) , \iota_\Phi  ( 0 , \xi ) ) \leq - c_0 |\xi|^2 \leq - c_1 | \delta_z |^2 .
\end{split}
\end{equation}
The left hand side of \eqref{eq:negL} equals  $ - 2 \langle ( \Im \psi_{zz}'' ) \delta_ z,
\delta_z \rangle $ and hence $ \Im \psi_{zz}'' > 0 $.
\end{proof}

\noindent
{\bf Remark.} The statement \eqref{eq63} means that $ T_0^* \mathbb C^n $ is
strictly negative with respect to $ T_{(x_0, \xi_0) } \Lambda_\Phi $.
Similarly, \eqref{eq:negL} means that the Lagrangian plane $ V $ in \eqref{eq64} is strictly negative
with respect to $ \mathbb R^{2n} $ (or simply strictly negative, \cite[Definition 21.5.5]{H3}).
For a detailed presentation of such concepts we refer to \cite[Chapter 11]{sam}, see also \cite{CoHiSj}.

\section{Analysis of the principal symbol}
\label{s:prince}

Here we essentially follow the arguments of~\cite[Section 3]{him}, specializing them to the setting of
Theorem \ref{t:2}. Let $q(y,\eta)$ be a real analytic function defined in a neighbourhood of $(y_0,\eta_0)\in T^*\mathbb R^2$, such that \begin{equation}
\label{eq64.1}
q(y_0,\eta_0) = 0.
\end{equation}
Assume also that
\begin{equation}
\label{eq65}
H_{{\rm Re}\, q}(y_0,\eta_0) \neq 0.
\end{equation}
Arguing as in~\cite[Theorem 21.3.6]{H3},~\cite{him},  using Darboux's theorem together with the implicit function theorem for holomorphic functions, we conclude that there exists a real analytic canonical transformation
\begin{equation}
\label{eq66}
\kappa: {\rm neigh}_{T^*\mathbb R^2}(y_0,\eta_0) \rightarrow {\rm neigh}_{T^*\mathbb R^2}(0,0) ,\quad \kappa(y_0,\eta_0) = (0,0),
\end{equation}
and a real analytic function $a$ defined in a neighbourhood of $(y_0,\eta_0)$, with $a(y_0,\eta_0)\neq 0$, such that
\begin{equation}
\label{eq67}
(aq)\circ \kappa^{-1} = \eta_1 + if(y,\eta_2).
\end{equation}
Here $f$ is real analytic and real valued in a neighbourhood of $(0,0)\in \mathbb R^2 \times \mathbb R$. We shall now strengthen the assumption (\ref{eq65}) by assuming that
\begin{equation}
\label{eq68}
H_{{\rm Re}\, q}(y_0,\eta_0), \quad H_{{\rm Im}\, q}(y_0,\eta_0) \quad \wrtext{are linearly independent},
\end{equation}
and we shall also assume that
\begin{equation}
\label{eq69}
\{q,\bar{q}\}(y_0,\eta_0) = -2i \{{\rm Re}\,q,{\rm Im}\,q\}(y_0,\eta_0) = 0.
\end{equation}
Writing $q = q_1 + iq_2$, $a = a_1 + ia_2$, with $q_j$, $a_j$ real, and observing that
\begin{equation}
\label{eq70}
H_{{\rm Re}(aq)}(y_0,\eta_0) = a_1(y_0,\eta_0) H_{q_1}(y_0,\eta_0) - a_2(y_0,\eta_0)H_{q_2}(y_0,\eta_0),
\end{equation}
\begin{equation}
\label{eq71}
H_{{\rm Im}(aq)}(y_0,\eta_0) = a_2(y_0,\eta_0) H_{q_1}(y_0,\eta_0) + a_1(y_0,\eta_0)H_{q_2}(y_0,\eta_0),
\end{equation}
we conclude that
\begin{equation}
\label{eq72}
H_{{\rm Re}\, (aq)}(y_0,\eta_0), \quad H_{{\rm Im}\, (aq)}(y_0,\eta_0) \quad \wrtext{are linearly independent}.
\end{equation}
We also have
\begin{equation}
\label{eq73}
\{aq,\overline{aq}\}(y_0,\eta_0) = \abs{a}^2 \{q,\bar{q}\}(y_0,\eta_0) =0.
\end{equation}
It then follows from (\ref{eq67}), (\ref{eq73}) that
\begin{equation}
\label{eq74}
\{\eta_1, f\}(0,0) = f'_{y_1}(0) = 0.
\end{equation}
Furthermore, (\ref{eq67}), (\ref{eq72}), (\ref{eq74}), and Jacobi's theorem, show that
\begin{equation}
\label{eq75}
H_{\eta_1}(0,0) = \partial_{y_1}, \ \ \ H_{f}(0,0) = f'_{\eta_2}(0)\partial_{y_2} - f'_{y_2}(0)\partial_{\eta_2}
\end{equation}
are linearly independent. The real valued real analytic function $(y_2,\eta_2)\mapsto f(0,y_2,\eta_2)$ has therefore a non-vanishing differential at $(0,0)$, and an application of Darboux's theorem allows us to conclude that there exists a real analytic canonical transformation
\begin{equation}
\label{eq76}
\begin{gathered}
 {\rm neigh}_{T^*\mathbb R^{2}} (0,0) \longrightarrow {\rm neigh}_{T^*\mathbb R^{2}}(0,0) \\
 (y_1,\eta_1;y_2,\eta_2) \longmapsto (\widetilde{y}_1,\widetilde{\eta}_1; \widetilde{y}_2,\widetilde{\eta}_2) =
(y_1,\eta_1; \widetilde{\kappa}(y_2,\eta_2)) , \ \  \widetilde \kappa ( 0 , 0 ) = ( 0 , 0 ) \end{gathered} \end{equation}
such that in these new coordinates,  $f(0,y_2,\eta_2) = \widetilde{\eta}_2$. Since
\begin{equation}
\label{eq77}
f(y,\eta_2) = f(0,y_2,\eta_2) + y_1 g(y,\eta_2),
\end{equation}
we can compose the symbol in (\ref{eq67}) with an additional real analytic canonical transformation of the form (\ref{eq76}), to obtain a reduction to a symbol of the form
\begin{equation}
\label{eq78}
\eta_1 + i\left(\eta_2 + y_1g(y,\eta_2)\right),
\end{equation}
where we know thanks to (\ref{eq74}) that $g(0) = 0$.

We summarize this in the following proposition:
\begin{prop}
\label{prop_3}
Let $q$ be a real analytic function in a neighbourhood of $(y_0,\eta_0)\in T^*\mathbb R^2$, such that $q(y_0,\eta_0) = 0$. Assume that
$H_{{\rm Re}\, q}(y_0,\eta_0)$, $ H_{{\rm Im}\, q}(y_0,\eta_0) $  are linearly independent, and
\begin{equation}
\label{eq80}
\{q,\bar{q}\}(y_0,\eta_0) = 0.
\end{equation}
There exists a real analytic canonical transformation
\begin{equation}
\label{eq81}
\kappa: {\rm neigh}_{T^*\mathbb R^2}(y_0,\eta_0) \rightarrow {\rm neigh}_{T^*\mathbb R^2}(0,0)  ,\quad \kappa(y_0,\eta_0) = (0,0),
\end{equation}
and a real analytic function $a$ defined in a neighbourhood of $(y_0,\eta_0)$, with $a(y_0,\eta_0)\neq 0$, such that
\begin{equation}
\label{eq82}
(aq)\circ \kappa^{-1} = {q_1}(y,\eta) := \eta_1 + i\left(\eta_2 + y_1g(y,\eta_2)\right).
\end{equation}
Here $g$ is real valued real analytic and 
$ g(0) = 0 $.
\end{prop}

We now add an assumption involving the second Poisson bracket:
\begin{equation}
\label{eq83}
\{q,\{q,\bar{q}\}\}(y_0,\eta_0) \neq 0,
\end{equation}
and seek a stronger analogue of Proposition \ref{prop_3} in this case.

A simple computation using (\ref{eq64.1}), (\ref{eq69}) gives then that
\begin{equation}
\label{eq84}
\{aq,\{aq,\overline{aq}\}\}(y_0,\eta_0) = \abs{a}^2 a\{q,\{q,\bar{q}\}\}(y_0,\eta_0) \neq 0,
\end{equation}
and therefore, in view of (\ref{eq82}), we have
\begin{equation}
\label{eq85}
\{q_1,\{q_1,\bar q_1\}\}(0,0) \neq 0.
\end{equation}
Here we have, in view of (\ref{eq82})
\begin{equation}
\label{eq86}
\{q_1,\bar q_1\}(y,\eta) = \tfrac{2}{i} \{\eta_1, \eta_2 + y_1 g\} = \tfrac{2}{i}\left(g(y,\eta_2) + y_1 g'_{y_1}(y,\eta_2 ) \right) ,
\end{equation}
and combining (\ref{eq86}) with the fact that
\begin{equation}
\label{eq87}
H_{q_1} = \partial_{y_1} + i\partial_{y_2} + {\mathcal O}((y,\eta))(\partial_{y},\partial_{\eta}),
\end{equation}
we get
\begin{equation}
\label{eq88}
\{q_1,\{q_1,\bar q_1\}\}(0,0) = \tfrac{2}{i}\left(2g'_{y_1}(0) + ig'_{y_2}(0)\right) \neq 0.
\end{equation}

The next step in the normal form construction is a reduction to the case when we have $g'_{y_1}(0) \neq 0$, $g'_{y_2}(0) = 0$ in (\ref{eq88}), and when carrying out this step we proceed as in~\cite[Lemma 3.3]{him}. Let us set $\lambda := 2g'_{y_1}(0) - ig'_{y_2}(0) \neq 0$, so that in view of (\ref{eq82}),
\begin{equation}
\label{eq89}
(\lambda aq)\circ \kappa^{-1} = \lambda q_1(y,\eta) =: q_2(y,\eta).
\end{equation}
Here (\ref{eq88}) gives that
\begin{equation}
\label{eq90}
\{q_2,\{q_2,\bar q_2\}\}(0,0) = \abs{\lambda}^2 \lambda \{q_1,\{q_1,\bar q_1\}\}(0,0) = -2\abs{\lambda}^4 i \in i\mathbb R \backslash \{0\}.
\end{equation}
Writing $\lambda = a + ib$, $a,b\in \mathbb R$, we get using (\ref{eq82}),
\begin{equation}
\label{eq91}
q_2(y,\eta) = (a+ib)\left(\eta_1 + i\eta_2 + iy_1g\right) = \left(a\eta_1 - b\eta_2 -by_1 g\right) + i  \left(b\eta_1 + a\eta_2 + ay_1g\right).
\end{equation}
We shall now simplify $q_2$ by means of a real linear canonical transformation. To this end, let
\begin{equation}
\label{eq92}
\kappa_0: T^*\mathbb R^2 \in (y,\eta) \mapsto ((C^t)^{-1}y,C\eta)\in T^*\mathbb R^2,
\end{equation}
where the invertible real $2\times 2$ matrix $C$ is given by
\begin{equation}
\label{eq93}
C = \begin{pmatrix}
a & -b \\\
b & \ a
\end{pmatrix}.
\end{equation}
We get using (\ref{eq91}), (\ref{eq92}),
\begin{equation}
\label{eq94}
q_2\left(\kappa_0^{-1}(y,\eta)\right) = \eta_1 + i\eta_2 + G(y,\eta), \quad G(y,\eta) = {\mathcal O}((y,\eta)^2),
\end{equation}
and therefore, incorporating the non-vanishing factor $\lambda$ into the function $a$ in (\ref{eq82}) and replacing the canonical transformation $\kappa$ in (\ref{eq81}) by $\kappa_0\circ \kappa$, we conclude that
\begin{equation}
\label{eq95}
(aq)\circ \kappa^{-1} = q_2(y,\eta) := \eta_1 + i\eta_2 + G(y,\eta), \quad \quad G(y,\eta) = {\mathcal O}((y,\eta)^2),
\end{equation}
where
\begin{equation}
\label{eq96}
\{q_2,\{q_2,\bar q_2\}\}(0,0) \in i\mathbb R \backslash \{0\}.
\end{equation}
Taking advantage of (\ref{eq96}), we can therefore proceed with the reduction of $q_2$ to a normal form, essentially by repeating the arguments above. We have
\begin{equation}
\label{eq97}
q_2(0,0) = 0,\quad \partial_{\eta_1} q_2(0,0) = 1,
\end{equation}
and using the implicit function theorem, we obtain the factorization,
\begin{equation}
\label{eq98}
q_2(y,\eta) = c_1 (y,\eta)(\eta_1 + r(y,\eta_2)),
\end{equation}
where $c_1 $ and $r$ are real analytic, with $r(0,0) = 0$, $c_1 (0,0) \neq 0$. Comparing the Taylor expansions of both sides of (\ref{eq98}) and using (\ref{eq95}), we conclude that
$ c_1 (0,0) = 1$.
We rewrite (\ref{eq98}) as follows: $ c_2 q_2 = \eta_1 + r(y,\eta_2)$ , $c_2 := c_1^{-1}$.
Using the Darboux theorem, applied to the system $y_1$, $\eta_1 + {\rm Re}\, r(y,\eta_2)$, satisfying
$$
\{\eta_1 + {\rm Re}\, r(y,\eta_2),y_1\} = 1,
$$
we next obtain a real analytic canonical transformation giving a reduction of $c_{ {2}} q_2$ to a real analytic function of the form $\eta_1 + if(y,\eta_2)$, where $f$ is real valued --- indeed, this is essentially a repetition of the arguments in the beginning of the discussion. Continuing in the same vein and repeating the arguments leading to Proposition \ref{prop_3} we conclude that there exists a real analytic canonical transformation
\begin{equation}
\label{eq101}
\kappa_1: {\rm neigh}_{T^*\mathbb R^2}(0,0) \rightarrow {\rm neigh}_{T^*\mathbb R^2}(0,0),\quad \kappa_1(0,0) = (0,0),
\end{equation}
such that
\begin{equation}
\label{eq102}
(c_{ {2}} q_2)\circ \kappa_1^{-1} = q_3(y,\eta) := \eta_1 + i\left(\eta_2 + y_1  g_3(y,\eta_2)\right),\quad g_3(0)= 0.
\end{equation}
Here we know, thanks to (\ref{eq96}) and the fact that $c_{ {2}} (0,0) = 1 $, that
\begin{equation}
\label{eq103}
\{c_{ {2}}q_2,\{c_{ {2}}q_2,\overline{c_{ {2}} q_2}\}\}(0,0) \in i\mathbb R \backslash \{0\},
\end{equation}
and it follows therefore, similarly to (\ref{eq88}), that
\begin{equation}
\label{eq104}
\{{q}_3,\{{q}_3,\bar {q}_3\}\}(0,0) =  - 2 i \left(2\partial_{y_1} g_3(0) + i\partial_{y_2} g_3(0)\right) \in i\mathbb R\backslash\{0\}.
\end{equation}
We obtain therefore that $\partial_{y_1} g_3(0) \neq 0$, $\partial_{y_2} g_3(0) = 0$.

Changing the notation for convenience (replacing $ q_3 $ by $ q_{ {0}}$ and $g_3 $ by $g $), we proved the  main result of this section:
\begin{prop}
\label{prop_4}
Let $q$ be a real analytic function in a neighbourhood of $(y_0,\eta_0)\in T^*\mathbb R^2$, such that $q(y_0,\eta_0) = 0$, and assume that
$ H_{{\rm Re}\, q}(y_0,\eta_0)$ and $ H_{{\rm Im}\, q}(y_0,\eta_0) $ are linearly independent,
and that
\begin{equation}
\label{eq106}
\{q,\bar{q}\}(y_0,\eta_0) = 0, \quad \{q,\{q,\bar{q}\}\}(y_0,\eta_0) \neq 0.
\end{equation}
Then there exists a real analytic canonical transformation
\begin{equation}
\label{eq107}
\kappa: {\rm neigh}_{T^*\mathbb R^2} (y_0,\eta_0) \rightarrow {\rm neigh}_{T^*\mathbb R^2} (0,0),\quad \kappa(y_0,\eta_0) = (0,0),
\end{equation}
and a real analytic function $a$ defined in a neighbourhood of $(0,0)$, with $a(0,0)\neq 0$, such that
\begin{equation}
\label{eq108}
q\circ \kappa^{-1} = a(y,\eta) q_{ {0}}(y,\eta) , \quad q_{ {0}}(y,\eta):= \eta_1 + i\left(\eta_2 + y_1g(y,\eta_2)\right),
\end{equation}
where $g$ is real valued real analytic satisfying
 $ g(0) = 0$, $g'_{y_1}(0) \neq 0$, and $g'_{y_2}(0) = 0$.
\end{prop}

\section{The complex eikonal equation and plurisubharmonic weights}
\label{s:eik}

Let $q(y,\eta)$ be a real analytic function defined in a neighbourhood of $(y_0,\eta_0)\in T^*\mathbb R^{ {2}}$,  and assume that $q(y_0,\eta_0) = 0$, $dq(y_0,\eta_0) \neq 0$. It follows from~\cite[Lemma 7.7]{sam} (we will provide a complete proof in our setting) that there exists a holomorphic function $\varphi(x,y)$ in a neighbourhood of $(x_0,y_0) \in \mathbb C^{ {2}} \times \mathbb R^{ {2}}$, for a suitable $x_0$ of the form $x_0 = (0,x_{02})$, satisfying $-\varphi '_y(x_0,y_0)=\eta _0\in {\mathbb R}^{ {2}}$,
\begin{equation}
\label{eq1}
\Im  \varphi''_{yy}(x_0,y_0)>0,\quad \det \varphi ''_{xy}(x_0,y_0)\ne 0,
\end{equation}
and such that the following complex eikonal equation holds:
\begin{equation}
\label{eq2}
\varphi'_{x_1}(x,y) = q(y,-\varphi'_y(x,y)).
\end{equation}
As in \eqref{eq54} we associate with $ \varphi $ a complex canonical transformation
$ \kappa_\varphi $. The eikonal equation \eqref{eq2} is equivalent to
\begin{equation}
\label{eq4}
q\circ \kappa^{-1}_\varphi  ( x, \xi) = \xi_1 .
\end{equation}
 We also associate to $ \varphi $ the strictly plurisubharmonic function $ \Phi$ as in
\eqref{eq6}.

For $ q $ with the special properties of \S \ref{s:prince} we want to construct a special solution
of \eqref{eq2} with $ \Phi $ having favourable properties. For that we follow the
approach of~\cite{him} (see also~\cite{Hi2}) with some simplification due to our special setting.

 Following Proposition \ref{prop_4}, let $a$ be real analytic in a neighbourhood of $(0,0)\in T^*\mathbb R^2$, $a(0)\neq 0$, and put
\begin{equation}
\label{eq110}
q_0(y,\eta) = \eta_1 + i(\eta_2 + y_1 g(y,\eta_2)),\quad (y,\eta) \in {\rm neigh}_{T^*\mathbb R^2} (0,0),
\end{equation}
where $g $ is real valued real analytic, and
\begin{equation}
\label{eq110.1}
g(0) = 0,\quad c:=g'_{y_1}(0) \neq 0, \quad g'_{y_2}(0) = 0.
\end{equation}
We shall be concerned with the model symbol $aq_0$ constructed in Proposition \ref{prop_4}, see (\ref{eq108}).

Taylor's formula and (\ref{eq110.1}) give
\begin{equation}
\label{eq110.2}
q_0(y,\eta) = \eta_1 + i\eta_2 + ic y_1^2 + i\alpha y_1 \eta_2 + iy_1 r(y,\eta_2),
\end{equation}
where $\alpha := g'_{\eta_2}(0)\in \mathbb R$ and
\begin{equation}
\label{eq110.3}
r(y,\eta_2) = \int_0^1 (1-t) g''_{(y,\eta_2),(y,\eta_2)}(ty, t\eta_2)(y,\eta_2)\cdot (y,\eta_2)\, dt = {\mathcal O}((y,\eta_2)^2).
\end{equation}
We also write
\begin{equation}
\label{eq110.31}
a(y,\eta) = a(0) + b(y,\eta),\quad b(y,\eta) = \int_0^{{1}} \nabla_{ y, \eta } a(ty,t\eta) \cdot (y,\eta)\,dt = {\mathcal O}((y,\eta)).
\end{equation}
Since the function $q$ in Proposition \ref{prop_4} can be multiplied by a non-vanishing  {constant} factor, we may assume in what follows that $a(0) = 1$.

We now introduce a rescaling parameter $\mu \in (0,1)$ and define
\begin{equation}
\label{eq110.4}
y = \mu \widetilde{y},\quad \eta = \mu^2 \widetilde{\eta}.
\end{equation}




Eventually, $ \mu$ will be taken sufficiently  {small} but fixed, that is, independent of $h$.

The transformation (\ref{eq110.4}) is not canonical as the symplectic form changes by the factor of $\mu^3$,
$ d \eta \wedge d y = \mu^3 d \widetilde \eta \wedge d \widetilde y $.
On the level of operators, this corresponds to a rescaling of the semiclassical parameter:
\begin{equation}
\label{eq110.5}
(aq_0)^w(y,hD_y) = (aq_0)^w(\mu \widetilde{y}, \mu^2 \widetilde{h} D_{\widetilde{y}}) = (\widetilde{a}\widetilde{q}_0)^w(\widetilde{y},\widetilde{h}D_{\widetilde{y}}),\quad \widetilde{h} := \mu^{-3} {h},
\end{equation}
where, in view of (\ref{eq110.2}), (\ref{eq110.31}),
\begin{equation}
\label{eq110.6}
\widetilde{q}_0(\widetilde{y}, \widetilde{\eta}) = q_0(\mu \widetilde{y},\mu^2 \widetilde{\eta}) = \mu^2\left(\widetilde{\eta}_1 + i\widetilde{\eta}_2 + ic \widetilde{y}_1^2\right) + i\mu^3 \alpha \widetilde{y}_1\widetilde{\eta}_2 + i\mu \widetilde{y}_1 r(\mu \widetilde{y},\mu^2\widetilde{\eta}_2),
\end{equation}
and
\begin{equation}
\label{eq110.61}
\widetilde{a}(\widetilde{y}, \widetilde{\eta}) = a(\mu \widetilde{y},\mu^2 \widetilde{\eta}) = 1 + b(\mu \widetilde{y},\mu^2\widetilde{\eta}).
\end{equation}
Dividing $\widetilde{a}\widetilde{q}_0$ by $\mu^2$ and dropping the tildes, we obtain the following normal form,
\begin{equation}
\label{eq110.7}
(aq_0)(y,\eta) = \left(1 + b(\mu y,\mu^2 \eta)\right)\left(\eta_1 + i\eta_2 + icy_1^2 + \mu r_{0,\mu}(y,\eta_2)\right),
\end{equation}
where
\begin{equation}
\label{eq110.8}
r_{0,\mu}(y,\eta_2) = i\alpha y_1 \eta_2 + iy_1 \int_0^1 (1-t) g''_{(y,\eta_2),(y,\eta_2)}(t\mu y, t\mu^2 \eta_2)(y,\mu \eta_2)\cdot (y,\mu \eta_2)\, dt,
\end{equation}
and therefore, we get
$ (aq_0)(y,\eta) = \eta_1 + i\eta_2 + icy_1^2 + \mu r_{\mu}(y,\eta)$, where
\begin{equation*}
\begin{split} & r_{\mu}(y,\eta) = r_{0,\mu}(y,\eta_2) \\
& \ \ + \left(\eta_1 + i\eta_2 + icy_1^2 + \mu r_{0,\mu}(y,\eta_2)\right){\int_0^1
\left(\nabla_{y,\eta} a\right)(t\mu y, t\mu^2\eta)\cdot (y,\mu\eta)\,dt}
\end{split}
\end{equation*}
is holomorphic in $(y,\eta) \in {\rm neigh}_{\mathbb C^2 \times \mathbb C^2}(0,0)$, with $C^{\infty}$ dependence on $\mu \in [0,1)$. We also notice that
$ r_{\mu}(y,\eta) = {\mathcal O}((y,\eta)^2)$,
uniformly in $\mu\in [0,1)$.

The Cauchy--Kowalevski theorem (see for instance \cite{N72}, \cite[Chapter 5]{Tre22}) applied to \eqref{eq2} with $ q = a q_0 $,
\begin{equation}
\label{eq111}
\begin{split}
\varphi'_{x_1}(x,y) & = (aq_0)(y,-\varphi'_y(x,y)) \\
& = - \varphi'_{y_1}(x,y) - i \varphi'_{y_2}(x,y) + icy_1^2 + \mu r_{\mu}(y,-\varphi'_{y}(x,y)),
\end{split}
\end{equation}
and the initial condition
\begin{equation}
\label{eq112}
\varphi(0,x_2,y) = \tfrac i 2 (x_2 -y_2)^2 + iy_1^2,
\end{equation}
has a unique holomorphic solution in a small fixed $ \mu$-independent neighbourhood of $(0,0)\in \mathbb C^2_x \times \mathbb C^2_y$. 
The Cauchy problem obtained by taking leading ($\mu$-independent) terms  only on the right hand side of
\eqref{eq111},
\begin{equation}
\label{eq113}
\psi'_{x_1}(x,y) = - \psi'_{y_1}(x,y) - i \psi'_{y_2}(x,y) + icy_1^2,
\end{equation}
with the same Cauchy data,
\begin{equation}
\label{eq114}
\psi(0,x_2,y) = \varphi(0,x_2,y) = \tfrac i 2 (x_2 -y_2)^2 + i y_1^2,
\end{equation}
can be solved exactly:
\begin{equation}
\label{eq115}
\psi(x,y) = \tfrac i 2 (x_2 - y_2 + ix_1)^2 + i(y_1 - x_1)^2 + \tfrac13{ic}\left(y_1^3 - (y_1 - x_1)^3\right).
\end{equation}
Combining (\ref{eq111}), (\ref{eq112}), (\ref{eq113}), and (\ref{eq114}), we then see (from the Cauchy--Kowalevski theorem \cite{N72}, \cite[Chapter 5]{Tre22}) that in the sense of holomorphic functions in a fixed neighbourhood of the origin in $\mathbb C^2_x \times \mathbb C^2_y$, we have 
\begin{equation}
\label{eq116}
\varphi(x,y) = \psi(x,y) + {\mathcal O}(\mu).
\end{equation}
In view of (\ref{eq110.8}), (\ref{eq111}), and (\ref{eq112}),  $\varphi(0) = 0$, $\varphi'_{x,y}(0) = 0$. The eikonal equation also shows that the ${\mathcal O}(\mu)$ term in (\ref{eq116}) vanishes to the third order at the origin. In particular, we have that
\begin{equation}
\label{eq116.1}
\varphi''_{yy}(0) = \begin{pmatrix}
\varphi''_{y_1y_1} & \varphi''_{y_1y_2}\\\
\varphi''_{y_2y_1}& \varphi''_{y_2y_2}
\end{pmatrix}(0) = \begin{pmatrix}
2i & 0\\\
0 & i
\end{pmatrix}
\end{equation}
has a positive definite imaginary part, and
\begin{equation}
\label{eq116.2}
\varphi''_{xy}(0) = \begin{pmatrix}
\varphi''_{x_1y_1} & \varphi''_{x_1y_2}\\\
\varphi''_{x_2y_1}& \varphi''_{x_2y_2}
\end{pmatrix}(0) = \begin{pmatrix}
-2i &  1 \\\
0 & -i
\end{pmatrix}
\end{equation}
is invertible, uniformly in $\mu > 0$.

\bigskip
\noindent
The corresponding weight function $\Phi$ is given by
\begin{equation}
\label{eq117}
\Phi(x) = -\Im \varphi(x,y(x)), \quad x\in {\rm neigh}_{\mathbb C^2} (0),
\end{equation}
where $y(x)\in {\rm neigh}_{\mathbb R^2} (0)$ is the unique point where the function ${\rm neigh}_{\mathbb R^2} (0) \ni y \mapsto -\Im \varphi(x,y)$ achieves its maximum. The function $y(x)$ depends real analytically on $x\in {\rm neigh}_{\mathbb C^2} (0) $ and we have $y(0) = 0$, so that $\Phi(0) = 0$. In what follows, we shall use the following consequence of the Cauchy-Riemann equations: the function $y(x) \in {\rm neigh}_{\mathbb R^2} (0)$ in (\ref{eq117}) is the unique point such that
\begin{equation}
\label{eq117.00.1}
\Im \,\left(\varphi'_y(x,y(x))\right) = 0.
\end{equation}
It follows then that
\[
\begin{split}
\frac{2}{i}\frac{\partial \Phi}{\partial x}(x) & = \partial_x ( \varphi ( x, y ( x ) ) - \overline{\varphi(x,y(x))}) \\
& = \varphi'_x(x,y(x)) +
\varphi'_y(x,y(x))\,\partial_x y(x) - \overline{\varphi'_y(x,y(x))\,\partial_{\overline{x}}y(x)} = \varphi'_x(x,y(x)),
\end{split} \]
and in particular, $\partial_x \Phi(0) = 0$. Here we have used (\ref{eq117.00.1}) and the fact that $y(x)$ is real.

\medskip
\noindent
It will be convenient for us to compute the third order Taylor expansion of the weight function $\Phi$ in (\ref{eq117}), regarding the $\mathcal O(\mu)$ term in (\ref{eq116}) as a small perturbation.
\begin{lemm}
\label{l:weight_Taylor}
We have for $x\in {\rm neigh}_{\mathbb C^2}(0)$ and $0 \leq \mu$ sufficiently small,
\begin{equation}
\label{eq117.0.1}
\Phi(x) = \tfrac{1}{2} (\Im x_2 + \Re \,x_1)^2 + (\Im x_1)^2 -\tfrac{1}{3} c (\Re x_1)^3 + \mathcal O(\abs{x_1}^4) + \mathcal O(\mu)\abs{x}^3.
\end{equation}
Here we recall that $c \in \mathbb R$ is non-vanishing.
\end{lemm}
\begin{proof}
We shall make use of (\ref{eq117}), (\ref{eq117.00.1}). Let us write, in view of (\ref{eq115}), (\ref{eq116}),
\begin{equation*}
\begin{split}
\varphi'_{y_1}(x,y) & = 2i(y_1 -x_1) + ic(y_1^2 - (y_1 -x_1)^2) + \mu\, \mathcal O((x,y)^2) \\
& =  2i(y_1 -x_1) + ic(2y_1 x_1 -x_1^2) + \mu\, \mathcal O((x,y)^2),\\
\varphi'_{y_2}(x,y) & = -i(x_2 - y_2 + ix_1) + \mu\, \mathcal O((x,y)^2).
\end{split}
\end{equation*}
Here we have also used the observation that the $\mathcal O(\mu)$ term in (\ref{eq116}) vanishes to the third order at the origin. We see therefore that (\ref{eq117.00.1}) holds precisely when
\begin{equation}
\label{eq117.4}
(2 + 2c\,\Re x_1) y_1 = 2\Re x_1 + c\,\Re (x_1^2) + \mu\, \mathcal O((x,y)^2),
\end{equation}
and
\begin{equation}
\label{eq117.5}
y_2 = \Re x_2 - \Im x_1 + \mu\, \mathcal O((x,y)^2).
\end{equation}
We get therefore, in view of (\ref{eq117.4}), (\ref{eq117.5}), and the implicit function theorem,
\begin{equation}
\label{eq117.6}
\begin{split}
y_1(x) & = \frac{\Re x_1}{1 + c\Re x_1} + \tfrac{1}{2} c \Re (x_1^2) + \mathcal O(\abs{x_1}^3) + \mathcal O(\mu)\abs{x}^2 \\
& = \Re x_1 - c (\Re x_1)^2 + \tfrac{1}{2} c \Re (x_1^2) + \mathcal O(\abs{x_1}^3) + \mathcal O(\mu)\abs{x}^2
\\ & = \Re x_1 - \tfrac{1}{2}c\abs{x_1}^2 + \mathcal O(\abs{x_1}^3) + \mathcal O(\mu)\abs{x}^2,
\end{split}
\end{equation}
\begin{equation}
\label{eq117.61}
y_2(x) = \Re x_2 - \Im x_1 + \mathcal O(\mu)\abs{x}^2.
\end{equation}
Using (\ref{eq117}), (\ref{eq117.6}), and (\ref{eq117.61}), we can now compute the weight. We first observe, using (\ref{eq117.61}), that
\begin{equation}
\label{eq117.7}
\begin{split}
-\Im \left(\tfrac12 {i} (x_2 - y_2(x) + ix_1)^2\right) & = \Im \left(\tfrac{i}{2} \left(\Im x_2 + \Re \,x_1 +
\mathcal O(\mu)\abs{x}^2\right)^2\right) \\
& = \tfrac{1}{2} (\Im x_2 + \Re \,x_1)^2 + \mathcal O(\mu)\abs{x}^3.
\end{split}
\end{equation}
Next we compute, using (\ref{eq117.6}), that
\begin{equation}
\label{eq117.8}
\begin{split}
-\Im  \left(i(y_1(x) - x_1)^2\right)
& = -\Re \,\left((y_1(x) - x_1)^2\right) \\
& = -\Re \left(\left(-i\Im x_1  -
\tfrac{c}{2} \abs{x_1}^2 + \mathcal O(\abs{x_1}^3) + \mathcal O(\mu)\abs{x}^2\right)^2\right)  \\
& = (\Im x_1)^2 + \mathcal O(\abs{x_1}^4) + \mathcal O(\mu)\abs{x}^3.
\end{split}
\end{equation}
Similarly,
\begin{equation}
\label{eq117.9}
\begin{split}
-\Im  \left(-\tfrac{i}{3}c (y_1(x) - x_1)^3\right) & = \Re \left(\tfrac{1}{3} c (y_1(x) -x_1)^3\right) \\
& = \tfrac{1}{3} c \Re \left(\left(-i\Im x_1  -
\tfrac{c}{2} \abs{x_1}^2 + \mathcal O(\abs{x_1}^3) + \mathcal O(\mu)\abs{x}^2\right)^3\right)
\\ & = \mathcal O(\abs{x_1}^4) + \mathcal O(\mu)\abs{x}^3,
\end{split}
\end{equation}
and finally,
\begin{equation}
\label{eq117.91}
\begin{split}
-\Im  \left(\tfrac{i}{3} cy_1(x)^3\right) & = -\tfrac{1}{3}c y_1(x)^3
 = -\tfrac{1}{3} c \left(\Re x_1 + \mathcal O(\abs{x_1}^2) + \mathcal O(\mu)\abs{x}^2\right)^3 \\
& = -\tfrac{1}{3} c (\Re x_1)^3 + \mathcal O(\abs{x_1}^4) + \mathcal O(\mu)\abs{x}^3.
\end{split}
\end{equation}
Combining (\ref{eq115}), (\ref{eq116}), (\ref{eq117}), (\ref{eq117.7}), (\ref{eq117.8}), (\ref{eq117.9}), and (\ref{eq117.91}), we obtain (\ref{eq117.0.1}), completing the proof.
\end{proof}

Let $r_0 > 0$ be small enough fixed, i.e. independent of $\mu$, so that the weight function $\Phi$ in (\ref{eq117.0.1}) is defined in a neighborhood of the closure of the open bidisc $D(0,2r_0) \times D(0,2r_0) \subset \mathbb C^2$. Here $D(0,2r_0) \subset \mathbb C$ is the open disc of radius $2r_0$, centered at the origin. Let $0 < \delta \leq 1$ and let us set
\begin{equation}
\label{eq118}
\Psi_{\delta}(x_2) = \inf_{x_1 \in D(0,2\delta r_0)} \Phi(x_1,x_2),\quad x_2 \in D(0,2\delta r_0).
\end{equation}

In order to apply a semiclassical analogue of~\cite[Theorem 7.9]{sam}, we need to make the following observation.
\begin{lemm}
\label{l:minorant}
Let $\Psi_{\delta}$ be given in {\rm (\ref{eq118})}, and for $ \zeta \in D(0,2\delta r_0)$,  let
\begin{equation}
\label{eq119}
\widetilde{\Psi}_{\delta}(\zeta) = \sup \left\{u(\zeta);\,\, u \,\,\wrtext{subharmonic and}\,\, u \leq \Psi_{\delta} \,\,\wrtext{on}\,\,D(0,2\delta r_0)\right\},
\end{equation}
be the largest subharmonic minorant of $\Psi_{\delta}$ in the disc $D(0,2\delta r_0)$. For each $\delta > 0$ small enough and for each $\mu > 0$ small enough, we have
\begin{equation}
\label{120}
\widetilde{\Psi}_{\delta}(0) < \Phi(0) = 0.
\end{equation}
\end{lemm}
\begin{proof}
In what follows, in order to fix the ideas, we shall assume that $c > 0$ in (\ref{eq117.0.1}). We observe first that, in view of (\ref{eq117.0.1}),
\begin{equation}
\label{eq121}
\Psi_{\delta}(x_2) \leq \Phi(-\Im x_2,x_2) = f(x_2) + \mathcal O(\abs{x_2}^4) + \mathcal O(\mu)\abs{x_2}^3, \quad x_2 \in D(0,2\delta r_0),
\end{equation}
where
$ f(x_2) := \frac{1}{3} c\,(\Im x_2)^3.$
Similarly to (\ref{eq119}), we introduce for $ \zeta \in D(0,2\delta r_0)$
\begin{equation}
\label{eq122}
U_{\delta}(\zeta) = \sup \left\{u(\zeta);\,\, u \,\,\wrtext{subharmonic and}\,\, u \leq f  \,\,\wrtext{on}\,\,D(0,2\delta r_0)\right\}.
\end{equation}

An application of~\cite[Theorem 2]{green} gives that the functions $\widetilde{\Psi}_{\delta}$, $U_{\delta}$, defined in (\ref{eq119}), (\ref{eq122}), respectively, are continuous subharmonic on $D(0,2\delta r_0)$, and we claim that
\begin{equation}
\label{eq123}
U_1(0) < 0.
\end{equation}
Indeed, the submean value property for the subharmonic function $U_1$ shows that
\begin{equation}
\label{eq124}
U_1(0) \leq \frac{1}{\pi r_0^2} \int\!\!\!\int_{D(0,r_0)} U_1(\zeta)\, L(d\zeta) < \frac{1}{\pi r_0^2} \int\!\!\!\int_{D(0,r_0)} f(\zeta)\, L(d\zeta) = 0.
\end{equation}
Here $L(d\zeta)$ is the Lebesgue measure in $\mathbb C$ and we have also used that the non-negative continuous function $f - U_1 \geq 0$ satisfies
\begin{equation}
\label{eq125}
\int\!\!\!\int_{D(0,r_0)} (f - U_1)\, L(d\zeta) > 0,
\end{equation}
since $U_1$ is subharmonic in $D(0,r_0)$ and $f$ is strictly superharmonic for $\Im \zeta < 0$. We also note that $U_{\delta}(0) = \delta^3 U_1(0) < 0$.

It follows from (\ref{eq121}) that
\begin{equation}
\label{eq125.1}
\Psi_{\delta}(\zeta) \leq f(\zeta) + C\delta^4 + C\mu\,\delta^3, \quad \abs{\zeta} < 2\delta r_0,
\end{equation}
for some constant $C>0$. Given $u$ subharmonic in $D(0,2\delta r_0)$ such that $u \leq \Psi_{\delta}$ on $D(0,2\delta r_0)$, we get therefore in view of (\ref{eq125.1}),
\begin{equation}
\label{eq125.2}
u(\zeta) - C\delta^4 - C\mu\,\delta^3 \leq f(\zeta), \quad \abs{\zeta} < 2\delta r_0.
\end{equation}
Here $u - C\delta^4 - C\mu\,\delta^3$ is subharmonic and therefore, recalling (\ref{eq122}), we get
\begin{equation}
\label{eq125.3}
u(\zeta) - C\delta^4 - C\mu\,\delta^3 \leq U_{\delta}(\zeta),\quad \abs{\zeta} < 2\delta r_0.
\end{equation}
It follows, in particular, using also (\ref{eq123}), that
\begin{equation}
\label{eq125.4}
\widetilde{\Psi}_{\delta}(0) \leq U_{\delta}(0) + C\delta^4 + C\mu\,\delta^3 = \delta^3 (U_1(0) + C\delta + C\mu) < 0,
\end{equation}
provided that $\delta > 0$ and $\mu > 0$ are small enough. The proof is complete.
\end{proof}

\bigskip
\noindent
The discussion in this section is summarized in the following proposition.
\begin{prop}
\label{p:phi}
Suppose that $ a $ and $ q_{0} $ are as in Proposition {\rm \ref{prop_4}}. There exists $c_0 > 0$ such that for each $\mu > 0$ sufficiently small, there exists $\varphi = \varphi_{\mu}(x,y)$, a holomorphic function defined in $\{(x,y)\in \mathbb C^2;\, \abs{x} < c_0 \mu, \abs{y} < c_0 \mu\}$, with the properties
\begin{equation}
\label{eq125.5}
\varphi(0,0) = 0,\quad \varphi'_{x,y}(0,0) = 0,\quad {\rm det}\, \varphi''_{xy}(0,0) \neq 0, \quad \Im \varphi''_{yy}(0,0) > 0,
\end{equation}
such that the associated complex symplectomorphism
\begin{equation}
\label{eq125.6}
\kappa_{\varphi}: (y,-\varphi'_y(x,y)) \mapsto (x,\varphi'_x(x,y))
\end{equation}
satisfies
\begin{equation}
\label{eq125.7}
(a q_0) \circ \kappa_{ \varphi }^{-1} = \xi_1.
\end{equation}
Associated to $\varphi$ is the corresponding weight function $\Phi = \Phi_\mu$, given in {\rm (\ref{eq117})}, defined for $\abs{x} \leq \mathcal O(\mu)$, enjoying the following property: let us set for $\delta \in (0,1]$,
\begin{equation}
\label{eq126}
\Psi_{\delta,\mu}(x_2) = \inf_{\abs{x_1} < c_1 \delta \mu} \Phi_{\mu}(x_1,x_2),\quad \abs{x_2} < c_1 \delta \mu,
\end{equation}
where $c_1 > 0$ is a constant depending only on $a$ and $q_0$. Then for every $\delta > 0$ small enough and every $\mu > 0$ small enough, the largest subharmonic minorant $\widetilde{\Psi}_{\delta,\mu}$ of $\Psi_{\delta,\mu}$ in the disk $D(0,c_1\delta \mu)$, defined as in {\rm (\ref{eq119})}, satisfies
\begin{equation}
\label{eq127}
\widetilde{\Psi}_{\delta,\mu}(0) < \Phi_{\mu}(0) = 0.
\end{equation}
\end{prop}
\begin{proof}
Let $ \varphi_1 $ be the phase function obtained by solving \eqref{eq111} with the initial condition \eqref{eq112}. We note that this construction was conducted in the rescaled coordinates \eqref{eq110.4} and the neighbourhoods in which it was valid  were independent of $\mu$
(this was stressed after \eqref{eq112}). Since we also rescaled $ h $ to $  \widetilde h = \mu^{-3} h $ (see \eqref{eq110.5}), it follows from
(\ref{eq110.6}), (\ref{eq110.61}), and (\ref{eq111}) that if we set
\begin{equation}
\label{eq128}
\varphi_{\mu}(x,y) := \mu^3 \varphi_1 \left({x}/{\mu},{y}/{\mu}\right),
\end{equation}
then $\varphi_{\mu}$ solves the eikonal equation in the original coordinates, in an $\mathcal O(\mu)$--neighborhood of the origin,
\begin{equation}
\label{eq129}
\partial_{x_1} \varphi_{\mu}(x,y) = (aq_0) (y, -\partial_y \varphi_{\mu}(x,y)).
\end{equation}
Here $q_0$ and $a$ are given in (\ref{eq110}) and (\ref{eq110.31}), respectively. Using (\ref{eq115}), (\ref{eq116}), (\ref{eq116.1}), (\ref{eq116.2}), (\ref{eq128}), and (\ref{eq129}), we conclude that (\ref{eq125.5}) and (\ref{eq125.7}) hold.

\medskip
\noindent
It follows from (\ref{eq117}), (\ref{eq128}) that the weight function associated to $\varphi_{\mu}$ is of the form
\begin{equation}
\label{eq130}
\Phi_{\mu}(x) = \underset{
\abs{y} \leq c_0 \mu } \sup \left(-\Im \varphi_{\mu}(x,y)\right) = \mu^3 \Phi\left({x}/{\mu}\right),
\end{equation}
where $\Phi$ is given in (\ref{eq117.0.1}), and therefore we get, in view of (\ref{eq118}), (\ref{eq126}), (\ref{eq130}), with $c_1 = 2r_0$,
\begin{equation}
\label{eq131}
\Psi_{\delta,\mu}(\zeta) = \mu^3 \Psi_{\delta}\left({\zeta}/{\mu}\right), \quad \abs{\zeta} < c_1 \delta \mu.
\end{equation}
The largest subharmonic minorant $\widetilde{\Psi}_{\delta,\mu}$ of $\Psi_{\delta,\mu}$ in the disk $D(0,c_1\delta \mu)$ satisfies therefore
\begin{equation}
\label{eqeq132}
\widetilde{\Psi}_{\delta,\mu}(\zeta) = \mu^3 \widetilde{\Psi}_{\delta}\left({\zeta}/{\mu}\right), \quad \abs{\zeta} < c_1 \delta \mu,
\end{equation}
where $\widetilde{\Psi}_{\delta}$ is defined in (\ref{eq119}), and (\ref{eq127}) follows from Lemma \ref{l:minorant}.
\end{proof}

\bigskip
\noindent
In our applications in the next section, we shall choose $\delta > 0$ and $\mu > 0$ sufficiently small fixed in (\ref{eq126}), so that the conclusions of Proposition \ref{p:phi} would hold.

\section{Proof of Theorem 2}
\label{s:pr2}
In the spirit of \cite{him}, we repeat the strategy of the proofs of \cite[Theorem 7.8, 7.9]{sam}.
Since those results are stated in the scalar case, we take care to show that (not unexpectedly)
lower order matricial terms in \eqref{eq:defP} do not affect the argument. The key is the special
solution of the eikonal equation \eqref{eq2} produced in \S \ref{s:eik}.

We recall from \S \ref{s:WF} that a family $ h \mapsto u ( h ) \in \mathscr D' ( U;\mathbb C^p ) $ is
$h$-tempered if for every $ \omega \Subset U $ there exists $ N $ such that
$ \| u \|_{ H^{-N} ( \omega ) } \leq \mathcal O (h^{-N }) $. In what follows, $ U \subset \mathbb R^2$ will be a fixed open set while the neighbourhoods
$ \neigh_{\mathbb C^2} ( \bullet ) $ may need to be very small depending on phase functions,
cut-offs, amplitudes but not on $ u $.

We first consider a general principally scalar system of semiclassical differential operators with analytic coefficients:
\begin{equation}
\label{eq:defPP}
\begin{gathered}
P ( y, h D_{y} ) = Q  + h R , \ \ Q = q ( y, hD_{y}) , \ \  R = ( R_{jk} ( y, h D_{y}) )_{ 1 \leq j , k \leq p } , \\
q ( y, \eta ) = \sum_{ |\alpha | \leq m } a_\alpha ( y ) \eta^\alpha , \ \
R_{ jk} ( y, \eta ) = \sum_{|\alpha | \leq m_{jk} }  r_{jk \alpha } ( y ) \eta^\alpha ,
\end{gathered}
\end{equation}
where $ a_\alpha $ and $ r_{jk \alpha } $ are real analytic in $ U $. (Here and later we abuse
the notation slightly and, for scalar operators $ A $, write $ A $ rather than $ A \otimes I_{ \mathbb C^p} $,
when considering the action of vector valued distributions.)

\begin{prop}
\label{p:transp}
Let $ P $ be given by \eqref{eq:defPP} and let $(y_0,\eta_0) \in T^*U$ be such that
\begin{equation}
\label{eq:pt}  q ( y_0 , \eta_0 ) = 0 , \ \  d q ( y_0, \eta_0 ) \neq 0 .
\end{equation}
Suppose that $ \psi ( x, y ) $ is holomorphic satisfying \eqref{eq1} and \eqref{eq2} for
$ ( x, y ) \in \neigh_{\mathbb C^{2} }  (x_0 ) \times \neigh_{\mathbb C^{2} } ( y_0 ) $, $ \eta_0 = - \psi_y'  (x_0, y_0 ) \in \mathbb R^{2}$, for a suitable $x_0 \in \mathbb C^2$. Then there exists a (matrix valued) elliptic classical analytic symbol $ a ( x, y ; h ) $ defined near $ ( x_0, y_0 ) $ and
$ \chi \in C^\infty_{\rm{c}} ( U ) $, $ \chi \equiv 1 $ near $ y_0 $, such that
\begin{equation}
\label{eq:defTchi}
T_h u ( x ) := \int_{\mathbb R^n } \chi ( y ) u ( y ) e^{ \frac i h \psi ( x, y ) } a ( x, y  ; h ) dy , \ \  \
 \ \  x \in \neigh_{\mathbb C^{2} } ( x_0 ) ,
\end{equation}
satisfies, for every {\em $h$-tempered} family $ h \mapsto u ( h ) \in \mathscr D'(U;\mathbb C^p)$,
\begin{equation}
\label{eq139}  | h D_{x_{1} } T_h u ( x ) - T_h P u ( x ) | \leq C e^{ ( \Phi ( x ) - \delta )/h} , \ \
x \in \neigh_{\mathbb C^{2} } ( x_0 ) ,
\end{equation}
where $ C $ depends on $ u $ but $ \delta > 0  $  and $ \neigh_{\mathbb C^{2}} ( x_0 ) $ do not.
Here, as in {\rm (\ref{eq6})}, we set
\begin{equation}
\label{eq139a}
\Phi(x) = \underset{y}{{\sup}}\left(-{\rm Im}\, \psi(x,y)\right),
\end{equation}
the supremum being taken over a small real neighbourhood of $y_0$.
\end{prop}

We remark that for any $h$-tempered family $ u = u ( h ) $, there exists $V = \neigh_{\mathbb C^2} ( x_0 )$ such that the holomorphic function $T_hu$ satisfies 
\begin{equation}
\label{eq:temp2FBI}
\forall\, \varepsilon > 0 \ \exists \, C_\varepsilon >0, \,
| T_h u ( x ) | \leq C_\varepsilon e^{ ( \Phi ( x ) + \varepsilon ) /h }, \ \  x \in V, \  h\in (0,h_0].
\end{equation}
The construction of an FBI transform such that (\ref{eq139}) holds is well known in the scalar case~\cite[Chapters 7,9]{sam} (see also \cite[{Theorem 2.9.2}]{HiS}), and our purpose here is to verify that the construction extends to principally scalar systems of the form (\ref{eq:defPP}).

\begin{proof}[Proof of Proposition \ref{p:transp}]
To keep  the notation simple, we assume that in \eqref{eq:defPP}, $ m = 2$, $ m_{jk} =1 $,  and $p=2$ (the proof can be easily modified for the general case).  Our purpose is to construct an elliptic classical analytic symbol of order $0$ in $h$, $a(x,y;h)$,
defined in a neighbourhood of $(x_0,y_0)$, taking values in $ \Hom( \mathbb C^2, \mathbb C^2)$, so that for $ T = T_h $ given by (\ref{eq:defTchi}), the equation \eqref{eq139} holds, for which we use the following shorthand
 \begin{equation}
\label{eq139s}
hD_{x_1}  Tu = T Pu \quad \wrtext{in}\,\, H_{\Phi,x_0},
\end{equation}
for an $ h$-tempered family $ u = u(h) \in \mathscr D' ( U; \mathbb C^2) $, see (\ref{eq:Sjsp0}), (\ref{eq:Sjeq}). 
Writing $Q_y = Q(y,hD_y)$, $R_y = R(y,hD_y)$, we have
\begin{multline}
\label{eq141}
\int e^{i\psi(x,y)/h} a(x,y;h) \chi(y) \left(Q_y  u(y)\right) dy
= \int Q_y^t (e^{i\psi(x,y)/h} a(x,y;h) \chi(y)) u(y)dy,
\end{multline}
and
\begin{equation}
\label{eq142}
\int e^{i\psi(x,y)/h} a(x,y;h) \chi(y) R_y u(y) dy
= \int (R^t_y (e^{i\psi(x,y)/h} \chi(y) a^{t} (x,y;h)))^{t}u(y)dy.
\end{equation}
Here $Q^t$ is the real transpose of $Q$ and $ R^t$ is defined as
\begin{equation}
\label{eq143}
R^t := \begin{pmatrix}
R_{11}^t & R_{21}^t \\\
R_{12}^t & R_{22}^t
\end{pmatrix},
\end{equation}
with $R_{jk}^t$ being the real transpose of $R_{jk}$ in (\ref{eq:defPP}). We are led therefore, in view of (\ref{eq:defPP}), (\ref{eq:defTchi}), (\ref{eq139s}), (\ref{eq141}), and (\ref{eq142}), to the following system of transport equations,
\begin{multline}
\label{eq144}
\left((hD_{x_1} + \psi'_{x_1}) \right)(a(x,y;h)) = \left(e^{-i\psi(x,y)/h} \circ Q^t_y \circ e^{i\psi(x,y)/h} \right)(a(x,y;h)) \\
+ h \left(\left( e^{-i\psi(x,y)/h}  \circ R^t_y \circ e^{i\psi(x,y)/h} \right)a^{t} (x,y;h)\right)^{t}
\end{multline}
The transpose of $ Q $ is given by
\begin{equation}
\label{eq145}
Q^t_y = q(y,-hD_y) + h\ell(y,hD_y) + h^2 b(y),
\end{equation}
where $\ell(y,hD_y)$ is a semiclassical first order differential operator with real analytic coefficients, and $b$ is a real analytic function. It follows that
\begin{equation}
\label{eq146}
\begin{split}
&  e^{-i\psi(x,y)/h} \circ Q^t_y \circ e^{i\psi(x,y)/h}   = q(y,-\psi'_y(x,y) - hD_y)
 \\ & \ \ \ \ \ \ \ \ \ \ \ \ \ \ \ \ \ \ \ + h\ell(y,\psi'_y(x,y) + hD_y) + h^2 b(y) \\
& \ \ \ \ \ \ \   = q(y,-\psi'_y(x,y)) - q'_{\eta}(y,-\psi'_y(x,y))\cdot hD_y + (h/2i ) {\rm tr}\,(q''_{\eta\eta}\psi''_{yy}(x,y)) \\
& \ \ \ \ \ \ \ \ \ \ \ \ \ \ \ \ + q_2(y,-hD_y)
+ h\ell(y,\psi'_y(x,y)) +h\ell'_{\eta}(y,\psi'_y(x,y))\cdot hD_y + h^2 b(y).
\end{split}
\end{equation}
Here, with the notation in (\ref{eq:defPP}), we have
$$
q_2(y,\eta) = \sum_{\abs{\alpha} =2} a_{\alpha}(y) \eta^{\alpha}.
$$
Combining (\ref{eq146}) with the eikonal equation  \eqref{eq2}, we obtain that
\begin{multline}
\label{eq147}
hD_{x_1} + \psi'_{x_1} - (e^{-i\psi(x,y)/h} \circ Q^t_y \circ e^{i\psi(x,y)/h}) \\
= hD_{x_1} + q'_{\eta}(y,-\psi'_y(x,y))\cdot hD_y + hf(x,y) + h^2 A(y,D_y),
\end{multline}
where $f(x,y)$ is a holomorphic function and $A(y,D_y)$ is a second order holomorphic differential operator, in a complex neighbourhood of $(x_0,y_0)$. We have next (using (\ref{eq:defPP}) with $ m_{jk} = 1$),
\begin{equation}
\label{eq148}
R_{jk}^t(y,hD_y) = R_{jk}(y,-hD_y) + hd_{jk}(y), \quad 1\leq j,k \leq 2,
\end{equation}
where the functions $d_{jk}$ are real analytic, and therefore,
\begin{multline}
\label{eq149}
e^{-i\psi(x,y)/h} \circ R_{jk}^t(y,hD_y) \circ e^{i\psi(x,y)/h} = R_{jk}(y,-\psi'_y(x,y) - hD_y) + hd_{jk}(y) \\
= R_{jk}(y,-\psi'_y(x,y)) - \partial_{\eta} R_{jk}(y,-\psi'_y(x,y))\cdot hD_y + hd_{jk}.
\end{multline}
It follows, in view of (\ref{eq143}), (\ref{eq149}), that
\begin{equation}
\label{eq150}
h\, e^{-i\psi(x,y)/h}  \circ R^t_y \circ e^{i\psi(x,y)/h}  = hM(x,y) + h^2B(x,y,D_y),
\end{equation}
where $M = (M_{jk})_{1\leq j,k\leq 2}$ is a holomorphic function in a complex neighbourhood of $(x_0,y_0)$, with values in $ \Hom ( \mathbb C^2,\mathbb C^2)$, and $B(x,y,D_y) = (B_{jk}(x,y,D_y))$ is a $2\times 2$ matrix of first order holomorphic differential operators.

Using (\ref{eq144}), (\ref{eq147}), (\ref{eq150}), and viewing the amplitude $a$ as a column vector in $\mathbb C^4$,
we rewrite (\ref{eq144}) as follows,
\begin{equation}
\label{eq151}
\left(hL  + h E(x,y) + h^2 C(x,y,D_x,D_y)\right) a(x,y;h) = 0.
\end{equation}
Here
\begin{equation}
\label{eq152}
L = L(x,y,\partial_x,\partial_y) = \partial_{x_1} + q'_{\eta}(y,-\psi'_y(x,y))\cdot \partial_y,
\end{equation}
$E(x,y) \in {\rm Hol}({\rm neigh}_{\mathbb C^4}(x_0,y_0); {\Hom}(
\mathbb C^{4} ,\mathbb C^{4}))$, and
$$ C(x,y,D_x,D_y) = (C_{jk}(x,y,D_x,D_y))_{1\leq j,k\leq 4 }, $$ is  a matrix of second order holomorphic differential operators. When analyzing (\ref{eq151}), we may assume, after a translation, that $x_0 = y_0 = 0$, and to simplify the notation, we shall write $z = ( x, y ) \in {\rm neigh}_{\mathbb C^{4}}(0)$,
$ x_1 = z_1 $, $ z = ( z_1, z') $.

\medskip
\noindent
The holomorphic vector field $L = L(z ,\partial_z)$ in (\ref{eq152}) is transversal to the complex hyperplane $H \subset \mathbb C^{4}$ given by $z_1 = 0$, and we may introduce therefore holomorphic flow out coordinates $ w = (w_1,\dots , w_{4} )$ in a neighbourhood of $0$, centered at $0$, such that the hyperplane $H$ is given by the equation $w_1 = 0$ and $L = \partial_{w_1}$ (see also~\cite[Lemma 2.1]{KrLiSa}). Passing to the flow out coordinates and changing $w $ to $z$, we may rewrite (\ref{eq151}) as follows,
\begin{equation}
\label{eq153}
h\partial_{z_1} a(z;h) + \left(h E(z) + h^2 C(z,D_z)\right) a(z;h) = 0.
\end{equation}
Here $E\in {\rm Hol}({\rm neigh}_{\mathbb C^4}(0); {\Hom}(\mathbb C^4,\mathbb C^4))$ and $C(z,D_z)$ is a $4\times 4$ matrix of second order holomorphic differential operators in a neighbourhood of $0\in \mathbb C^{4}$. We shall solve (\ref{eq153}) demanding that $a(0,z';h) = b(z';h)$, where $b$ is a $\mathbb C^4$--valued classical analytic symbol in a neighbourhood of $0\in \mathbb C^{3} $, and as explained in ~\cite[Section 2.2]{KrLiSa}, when doing so it suffices to solve the initial value problem
\begin{equation}
\label{eq154}
\left\{ \begin{array}{ll}
h\partial_{z_1} a(z;h) + \left(h E(z) + h^2 C(z,D_z)\right) a(z;h) = hv(z;h), \\
a(0,z';h) = 0,
\end{array} \right.
\end{equation}
where $v(z;h)$ is a classical analytic symbol in a neighbourhood of $0\in \mathbb C^{4}$, with values in $\mathbb C^4$. To eliminate the matrix $E(z)$ from the transport equations (\ref{eq154}),
we introduce a fundamental matrix, that is an invertible  $F\in {\rm Hol}({\rm neigh}_{\mathbb C^4}(0); {\Hom}(\mathbb C^4,\mathbb C^4))$
satisfying
\begin{equation}
\label{eq156}
\partial_{z_1} F(z) + E(z) F(z) = 0.
\end{equation}
Looking for a solution to (\ref{eq154}) of the form $a(z;h) = F(z) \widetilde{a}(z;h)$, we see that $\widetilde{a}$ should satisfy
\begin{equation}
\label{eq157}
\left\{ \begin{array}{ll}
h\partial_{z_1} \widetilde{a}(z;h) + h^2 F^{-1}(z) \circ C(z,D_z)\circ F(z)\, \widetilde{a}(z;h) = hF^{-1}(z)v(z;h), \\
\widetilde{a}(0,z';h) = 0.
\end{array} \right.
\end{equation}
It follows therefore that when solving (\ref{eq154}), we may assume that $E(z) = 0$.

\medskip
\noindent
The analysis of (\ref{eq154}), when $E(z) = 0$, proceeds by means of the method of "nested neighbourhoods", developed in~\cite[Chapter 9]{sam} (see also \cite[\S 2.8]{HiS}) in the scalar case. An extension to the present matrix valued case is straightforward, and the following discussion is given for the completeness and convenience of the reader only -- see also~\cite{KrLiSa},~\cite{RZ}. Let
$ \Omega_0 = \{z\in \mathbb C^{4}; \abs{z_1} + \abs{z'} < r\}$,
where $r>0$ is small enough so that $\overline{\Omega_0}$ is a compact subset of the domain of definition $\Omega \subset \mathbb C^{4}$ of $v$ and $C(z,D_z)$ in (\ref{eq154}). We set
\begin{equation}
\label{eq159}
\Omega_t = \{z\in \mathbb C^{4}; \abs{z_1} + \abs{z'} < r-t\}, \quad 0\leq t < r.
\end{equation}

\medskip
\noindent
Given $\rho > 0$, we say that $a\in {\mathcal A}_{\rho}$, if $a(z;h) = \sum_{k=0}^{\infty} a_k(z)h^k$, $a_k \in {\rm Hol}(\Omega;\mathbb C^{4})$, is such that for all $t\in (0,r)$, we have
\begin{equation}
\label{eq160}
\underset{\Omega_t}{{\rm sup}} \abs{a_k} \leq {f(a,k)}k^k t^{-k} , \quad k\geq 0, 
\end{equation}
where $f(a,k)$ is the best constant for which (\ref{eq160}) holds, and 
\begin{equation}
\label{eq161}
 \| a \|_{\rho} := \sum_{k=0}^{\infty} f(a,k) \rho^k < \infty.
\end{equation}

\medskip
\noindent
Let
$ (\partial_{z_1}^{-1}a)(z_1,z') = \int_0^{z_1} a(y_1,z')\, dy_1$.
We then have the following well known result, see~\cite[Theorem 9.3]{sam},~\cite[Lemma 5.5]{RZ},~\cite[Lemma 2.2]{KrLiSa}.
\begin{lemm}
\label{prop_inverse}
For $a\in {\mathcal A}_{\rho}$ of the form $a(z;h) = \sum_{k=2}^{\infty} a_k(z) h^k$, and $b = (h\partial_{z_1})^{-1}a$,
 we have
\begin{equation}
\label{eq163}
\| b \|_{\rho} \leq  \frac{ 2e } \rho \| a \|_{\rho}.
\end{equation}
\end{lemm}
\begin{proof}
We write
$ b = \sum_{k=2}^{\infty} h^{k-1} \partial_{z_1}^{-1} a_k = \sum_{k=1}^{\infty} h^k b_k$,
where
\begin{equation}
\label{eq164}
b_k(z) = (\partial_{z_1}^{-1} a_{k+1})(z) = \int_0^{z_1} a_{k+1}(y_1,z')\,dy_1 = z_1 \int_0^1 a_{k+1}(\sigma z_1,z')\, d\sigma.
\end{equation}
It follows from (\ref{eq159}) that if $z\in \Omega_t$, we have
$
(\sigma z_1,z') \in \Omega_{t + (1-\sigma)\abs{z_1}}$, $ 0\leq \sigma \leq 1$.
Using this and (\ref{eq160}), we obtain for $z\in \Omega_t$,
\begin{equation}
\label{eq166}
\begin{split}
\abs{b_k(z)} & \leq \abs{z_1} f(a,k+1) (k+1)^{k+1} \int_0^1 \frac{d\sigma}{(t + (1-\sigma)\abs{z_1})^{k+1}}
 \\ &
= f(a,k+1) (k+1)^{k+1}
\int_0^{\abs{z_1}} \frac{d\sigma}{(t + \sigma)^{k+1}} \\
& \leq f(a,k+1) (k+1)^{k+1} \int_t^{\infty} \frac{d\sigma}{\sigma^{k+1}}
= f(a,k+1) \frac{(k+1)^{k+1}}{k t^k},\quad k\geq 1.
\end{split}
\end{equation}
Thus, for $0 < t < r$,
\begin{equation}
\label{eq167}
\underset{\Omega_t}{{\rm sup}} \abs{b_k} \leq \frac{2e f(a,k+1)}{t^k} k^k, \quad k\geq 1,
\end{equation}
and therefore, $f(b,k) \leq 2e f(a,k+1)$, $k=1,2,\dots$, implying that
\begin{equation}
\| b \|_{\rho} = \sum_{k=1}^{\infty} f(b,k) \rho^k \leq \sum_{k=1}^{\infty} 2e f(a,k+1) \rho^k = \frac{2e}{\rho} \| a \|_{\rho},
\end{equation}
which gives \eqref{eq163}.
\end{proof}

\medskip
\noindent
Applying $(h\partial_{z_1})^{-1}$ to (\ref{eq154}), we get, recalling that $E(z) = 0$,
\begin{equation}
\label{eq168}
a_j(z) + \sum_{k=1}^4 (h\partial_{z_1})^{-1} h^2C_{jk}(z,D_z)a_k(z) =  \partial_{z_1}^{-1}v_j(z),\quad 1\leq j \leq 4.
\end{equation}
 Next, we have the following result (see~\cite[Lemma 2.3]{KrLiSa}):
\begin{lemm}
\label{Neumann}
Let $a\in {\mathcal A}_{\rho}$ be scalar valued and let $Q = Q(z,D_z)$ be a second order holomorphic differential operator in $\Omega$. Then $(h\partial_{z_1})^{-1} h^2 Q(z,D_z)a\in {\mathcal A}_{\rho}$, with
\begin{equation}
\label{eq169}
\norm{(h\partial_{z_1})^{-1} h^2 Q(z,D_z)a}_{\rho} \leq {\mathcal O}(\rho)\| a \|_{\rho}.
\end{equation}
\end{lemm}
\begin{proof}
Writing $a(z;h) = \sum_{k=0}^{\infty} a_k(z) h^k$, we obtain that
$$
h^2 Q(z,D_z) a = \sum_{k=2}^{\infty} h^{k} Q(z,D_z) a_{k-2}.
$$
For $0 < s < t < r$, we get using the Cauchy estimates,
\begin{equation}
\label{eq170}
\begin{split}
\underset{\Omega_t}{{\rm sup}} \abs{Q(z,D_z)a_{k-2}} & \leq \frac{C}{(t-s)^2}\, \underset{\Omega_s}{{\rm sup}} \abs{a_{k-2}}
\\ & \leq \frac{C}{(t-s)^2} \frac{f(a,k-2)}{s^{k-2}}(k-2)^{k-2},\quad k\geq 2.
\end{split}
\end{equation}
Taking $ s = (k-2)t/k < t$, we get using (\ref{eq170}),
$$
\underset{\Omega_t}{{\rm sup}} \abs{Q(z,D_z)a_{k-2}} \leq \frac{C f(a,k-2)}{t^k} k^k, \ \ k \geq 3.
$$
Therefore, in view of (\ref{eq160}),
$
f(Q(z,D_z)a_{k-2},k) \leq C f(a,k-2),
$
and  definition (\ref{eq161}) gives
\begin{equation}
\label{eq171}
\norm{h^2 Q(z,D_z)a}_{\rho} \leq C \sum_{k=2}^{\infty} \rho^k f(a,k-2) \leq {\mathcal O}(\rho^2) \| a \|_{\rho}.
\end{equation}
Combining Lemma \ref{prop_inverse} with (\ref{eq171}), we obtain that
$$
\norm{(h\partial_{z_1})^{-1} h^2 Q(z,D_z)a}_{\rho} \leq C \rho^{-1} \norm{h^2 Q(z,D_z)a}_{\rho} \leq {\mathcal O}(\rho) \| a \|_{\rho},
$$
establishing (\ref{eq169}).
\end{proof}

Rewriting (\ref{eq168}) in the form
\begin{equation}
\label{eq172}
(1 + L)a = \partial_{z_1}^{-1}v, \ \ \
(La)_j := \sum_{k=1}^4 (h\partial_{z_1})^{-1} h^2C_{jk}(z,D_z)a_k(z), \quad 1\leq j \leq 4,
\end{equation}
we conclude using Lemma \ref{Neumann} that $\norm{La}_{\rho} \leq {\mathcal O}(\rho) \| a \|_{\rho}$, and therefore, for $\rho > 0$ small enough, the equation (\ref{eq172}) has a unique solution $a$ such that $\| a \|_{\rho} < \infty$. Thus, $a$ is a classical analytic symbol in a neighbourhood of the origin. Coming back to (\ref{eq153}) and demanding that $a|_{z_1 = 0}$ should be an elliptic classical analytic symbol near the origin in $\mathbb C^3$, we conclude that the classical analytic symbol $a(z;h)$ is elliptic. This completes the construction of a matrix valued FBI transform of the form (\ref{eq:defTchi}), such that (\ref{eq139s}) holds. The proof of Proposition \ref{p:transp} is complete.
\end{proof}

We can now prove a microlocal version of Theorem \ref{t:2}. When $ u $ is independent
of $ h$, $ \WFh (u) \setminus 0 $ (here $ 0 $ denotes the zero section in $ T^* U$)
is equal to the standard analytic wave front set $ \WF_{\rm a} (u) $ - see \cite[\S 8.4,9.6]{H1}, \cite[Chapter 6]{sam}.
However, the essential  aspect here is $ h$-dependence.

\begin{theo}
\label{t:3}
Suppose that $ P $ is given by \eqref{eq:defPP} and that  $ u \in \mathscr D' ( U; \mathbb C^2 ) $ is an
$ h$-tempered $ \mathbb C^2$-valued distribution. If at some point $ \rho \in T^* U $ we have
\begin{equation}
\label{eq:t31}  q ( \rho ) = \{ q, \bar q \} ( \rho ) =0 , \ \   \{ q, \{ q,  \bar q \}\} (\rho )  \neq 0 ,
\ \ H_{q} ( \rho)  \not{\! \parallel} \, \, H_{\bar q } (\rho ) ,
 \ \ \rho \notin \WFh ( P u ),  \end{equation}
then
$ \rho \notin \WFh ( u )$.
\end{theo}
\begin{proof}
We put $ \rho = ( y_0 , \eta_0 ) $. The first step of the proof is to construct a suitable holomorphic function $ \psi $ for which \eqref{eq1}, \eqref{eq2} hold, with $ (x_0, \xi_0 ) = ( 0, 0) $, $ - \psi'_y (0 , y_0 ) = \eta_0 $.

{We first use Proposition \ref{prop_4} to obtain, in the notation of that proposition, a real analytic local canonical transformation $\kappa$ such that $ q \circ \kappa^{-1} = a q_0 $. We then use Proposition \ref{p:phi}  to obtain $ \varphi  = \varphi_\mu $, for $ \mu > 0$ sufficiently small but fixed, such that we have locally,
\begin{equation}
\label{eq:aq0}   ( a q_0) \circ \kappa_\varphi^{-1} = \xi_1 .\end{equation} 
Throughout the following discussion, the parameter $\mu>0$ will be kept fixed and the dependence on $\mu$ will not be indicated explicitly. With \eqref{eq:aq0} in place,  Proposition \ref{p:phi2psi}  then shows that there exists a holomorphic function $\psi(x,y)$ defined near $(0,y_0)\in \mathbb C^2 \times \mathbb R^2$, satisfying
$$
- \psi'_y (0 , y_0 ) = \eta_0, \quad {\rm det}\, \psi''_{xy}(0,y_0) \neq 0, \quad {\rm Im}\, \psi''_{yy}(0,y_0) > 0, 
\quad \kappa_\varphi \circ \kappa = \kappa_\psi .
$$
 Hence
\[  q \circ \kappa_{\psi}^{-1} = q \circ \kappa^{-1} \circ \kappa_\varphi^{-1} =
( aq_0 ) \circ \kappa_\varphi^{-1} = \xi_1 . \]
This shows that the eikonal equation \eqref{eq2} holds for $ \psi$. If $ \Psi $ is the weight function 
associated to $ \psi $ as in \eqref{eq6} then, in the notation of \eqref{eq6},
\[ \begin{split} \Lambda_\Psi & = \kappa_\psi ( \neigh_{T^* \mathbb R^2 } ( y_0, \eta_0 ) ) =
\kappa_\varphi \circ \kappa ( \neigh_{T^* \mathbb R^2 } ( y_0, \eta_0 ) )  \\
& =
\kappa_\varphi ( \neigh_{ T^* \mathbb R^2 } ( 0 , 0 ) ) = \Lambda_\Phi . \end{split}  \]
Since $ \Lambda_\Psi $ determines $ \Psi $ up to an additive constant, we can
choose $ \psi $ so that $ \Psi = \Phi$ in a neighbourhood of $0\in \mathbb C^2$.} 

{We now apply Proposition \ref{p:transp} to $ P $ and conclude that there exist $\delta_1 >0$, $\varepsilon_0 > 0$, such that 
\eqref{eq139} holds with $ x_0 = 0$:
\begin{equation}
\label{eq139n}  h D_{x_1} T_h u ( x ) = T_h (P u) ( x )  + \mathcal O ( e^{ ( \Phi ( x ) - \delta_1 )/h} ) , \ \ \ \ 
| x| < \varepsilon_0  .
\end{equation}
Since $ \rho \notin \WFh ( P u ) $, Proposition \ref{p:FBI} shows that
$ T_h ( P u ) ( x ) = \mathcal O ( e^{ ( \Phi ( x ) - \delta_2 )/h} ) $, $ |x| < \varepsilon_0 $, for some $ \delta_2 > 0 $, and therefore  (\ref{eq139n}) gives  \begin{equation}
\label{eq_SjCh7.1}
h D_{x_1} T_h u ( x ) = \mathcal O ( e^{ ( \Phi ( x ) - \delta_3 )/h} ) , \ \ \ \
x=(x_1,x_2)\in \mathbb C^2, \ \ 
|x_1| < \varepsilon_1,\,\, |x_2| < \varepsilon_1, 
\end{equation}
for some $\delta_3 > 0$, $\varepsilon_1 > 0$. Since  $\Phi(0) = 0$,  we can find $0 < \eta \leq 1$ such that
\begin{equation}
\label{eq_SjCh7.2}
\abs{\Phi(x)} < \tfrac14{\delta_3},\quad  |x_1| < \eta\,\varepsilon_1,\,\, |x_2| < \eta\,\varepsilon_1.
\end{equation}
Hence, (\ref{eq_SjCh7.1}) and (\ref{eq_SjCh7.2}), and integration in $ x_1 $ give
\begin{equation}
\label{eq:02x}    
T_h u ( x_1, x_2 ) = T_h u ( y_1, x_2) + \mathcal O (1) e^{ -\delta_3/2h}, \ \  |x_1| < \eta\,\varepsilon_1,\,\, |y_1| < \eta\,\varepsilon_1,\,\,
|x_2| < \eta\,\varepsilon_1.
\end{equation}
Since $ u $ is assumed to be $h$-tempered,  \eqref{eq:temp2FBI} shows that for every
$ \varepsilon > 0 $, we have 
\begin{equation}
\label{eq:0y}
| T_h u ( y_1 , x_2 ) | \leq C_\varepsilon e^{ ( \Phi ( y_1 , x_2) + \varepsilon ) / h } , \ \ |y_1| < \eta\,\varepsilon_1,\,\,
|x_2| < \eta\,\varepsilon_1.
\end{equation}
Combining \eqref{eq:02x} and \eqref{eq:0y} we obtain for every $\varepsilon >0$,  
\begin{equation}
\label{eq_SjCh7.3}
| T_h u ( x )|  \leq C_\varepsilon e^{(\Psi_{\eta}(x_2) + \varepsilon) / h } + C e^{ -\delta_3/2h}, \quad |x_1| < \eta\,\varepsilon_1,\,\,|x_2| < \eta\,\varepsilon_1,
\end{equation}
where 
$$
\Psi_{\eta}(x_2) = \inf_{ |y_1| < \eta\,\varepsilon_1} \Phi ( y_1 , x_2 ).
$$
But then (\ref{eq_SjCh7.2}) and (\ref{eq_SjCh7.3}) show that
\begin{equation}
\label{eq_SjCh7.4}
| T_h u ( x )|  \leq C_\varepsilon e^{(\Psi_{\eta}(x_2) + \varepsilon ) / h }, \quad |x_1| < \eta\,\varepsilon_1,\,\,|x_2| < \eta\,\varepsilon_1,
\end{equation}
for every $ \varepsilon > 0 $.
Fixing a sufficiently small  $\eta > 0$  and using \eqref{eq126} and \eqref{eq127}, as described at the end of \S \ref{s:int} (see \eqref{eq:kash})  we see, in view of (\ref{eq:02x}) and (\ref{eq_SjCh7.4}), that $\abs{T_h u ( x )} \leq C e^{ - \delta/ h } $ for $ x \in \neigh_{\mathbb C^2}
( 0 ) $ and some $\delta > 0$. In view of Proposition \ref{p:FBI} this concludes the proof.}
\end{proof}

To deduce Theorem \ref{t:2} from Theorem \ref{t:3} we need the following result, stated for operators on $\mathbb R^n$, for all $n$. 
\begin{prop}
\label{p:global}
Assume that $ P $ is given by \eqref{eq:defPP} and that we have 
\begin{equation}
\label{eq:elli}
| q ( x, \xi ) | \geq C |\xi|^m - C, \quad (x,\xi) \in T^*U,\ \ m_{ij} \leq m .
\end{equation}
If $ P u = 0 $ near $ x_0 \in \mathbb R^n$, $  u $ is $h$-tempered and
\begin{equation}
\label{eq:WFass}  \WFh ( u ) \cap \, q^{-1} ( 0 ) \cap \pi^{-1} ( x_0 ) = \emptyset ,
\end{equation}
then there exists a neighbourhood $ \Omega $ of $ x_0 $ and $ C_0 , c_0   > 0 $ such that
\begin{equation}
\label{eq:expdec1}
      | \partial^\beta u ( x ) | \leq  C_0 ( |\beta | C_0 )^{|\beta|} e^{ - c_0 / h} , \ \ \ x \in \Omega, \ \
      \beta \in \mathbb N^n .
\end{equation}
\end{prop}

\noindent
{\bf Remark.} We prove this general result using somewhat advanced methods developed
in \cite{gaz1},\cite{gaz2} and based on \cite{HS} and \cite{Sj96}. For the concrete application
in Theorem \ref{t:1} we could use the fact that the coefficients of $ D ( \alpha ) $ are globally
analytic (in fact, entire in $ \mathbb C^2 $) and apply a small modification of
 \cite[Theorem 4.1.5]{mart} proved using
methods well explained in that text.

Let $u \in \mathscr D' ( U_0 )$, where $U_0 \subset \mathbb R^n$ is an open neighbourhood of $x_0$. To handle estimates away from $ q^{-1} ( 0 ) $ we will use a non-holomorphic FBI transform adapted to the study of the behaviour as $ |\xi | \to \infty $:
\begin{equation}
\label{eq:nFBI}
 T_{\Lambda_0}  u ( x, \xi )  :=  h^{-\frac{3n}{4}}\int_{\RR^n}
e^{ \frac i h ( (x-y)\cdot \xi +\frac{i}{2}\langle \xi\rangle (x-y)^2)}
\langle \xi\rangle^{\frac{n}{4}} \chi( y ) u(y)dy, 
\end{equation}
where $ \chi \in C^\infty_{\rm{c}} ( U_0 ) $, $ \chi ( x ) =1 $ in $ \neigh_{\mathbb R^n} ( x_0 ) $.
Here, to be consistent with the notation below, $ \Lambda_0 = T^* \mathbb R^n $.

In compact sets, the decay of  $ T_{\Lambda_0 }  $ and the FBI transform from \eqref{eq:WFa},
$ T_h $, are equivalent. The only reason this is not a special case of Proposition \ref{p:FBI} comes from the
fact that $ T_{\Lambda_0} $ is not a holomorphic FBI transform.

\begin{lemm}
\label{l:T2T}
Suppose that $ u = u ( h ) \in \mathscr D' ( U_0 ) $ is an $h$-tempered family
of distributions,
 $ T_h u $ is given in
\eqref{eq:WFa} and $  T_{\Lambda_0} u $ is defined by \eqref{eq:nFBI}. If $ U $ is a neighbourhood of  $ z_0 = x_0 - i \xi_0  \in \mathbb C^n $ such
that for some $ \delta > 0 $ and $0 < h \leq h_0$, 
\begin{equation}
\label{eq:T2T} | T_h u ( z ) | \leq  C e^{(\Phi( z )  - \delta ) /h } , \ \  z \in U , \ \ \ \Phi ( z ) := \tfrac12 |\Im z|^2,   \end{equation}
then there exist $ V \Subset \mathbb C^{2n} $, an open complex neighbourhood of $ ( x_0, \xi_0 )
\in \mathbb R^{2n}  $, 
and $ \delta' > 0 ,  C'> 0 $ such that
\begin{equation}
\label{eq:T2T2}  |T_{\Lambda_0} u ( x, \xi ) | \leq C' e^{ - \delta'/h } , \ \ ( x, \xi ) \in V . \end{equation}
\end{lemm}
\begin{proof}
We write $ T_{\Lambda_0 } \chi u = c_n h^{-\frac{3n}2} T_{\Lambda_0} T_h^* T_h \chi u $
(here we abuse the notation slightly and define $ T_h $ and $ T_{\Lambda_0} $
 without the cut-off $ \chi $)  and describe the operator
$ T_{\Lambda_0} T_h^*$ first. Here the adjoint of $T_h$ is taken with respect to
$ L^2 ( \mathbb C^n , e^{ -2 \Phi( z )  /h} L(dz ) ) $, $ c_n \neq 0 $ -- see \cite[Theorem 1.3.3]{HiS}.

{The Schwartz kernel of the composition $T_{\Lambda_0} T_h^*$ takes the form 
\begin{multline}
\label{eq:defK}  
K ( x, \xi, z ) =  h^{-\frac {3 n}  4 }\langle \xi \rangle^{\frac n4} e^{-2\Phi(z)/h} \int_{\mathbb R^n }
e^{ \frac i h ( x - y )\cdot \xi - \frac 1 { 2 h } \langle \xi \rangle ( x - y)^2 - \frac 1 { 2 h } ( \overline{z} - y )^2 } dy \\
= h^{-\frac {3 n}  4 }\langle \xi \rangle^{\frac n4} e^{\frac{i}{h}\left(x\cdot \xi + \Re z\cdot \Im\, z\right)} e^{-\Phi(z)/h} 
\int_{\mathbb R^n } e^{-iy\cdot (\xi + \Im z)/h} e^{-\frac 1 { 2 h } \langle \xi \rangle ( x - y)^2 - \frac 1 { 2 h } ( \Re z - y )^2 } dy.
\end{multline} 
Following the calculations in the proof of \cite[Proposition 3.2.5]{mart}, we obtain
\begin{equation}
\label{eq_mart1}
\langle \xi \rangle ( x - y)^2 + ( \Re z - y )^2 = (1 + \langle \xi \rangle) \left(y - \frac{\langle \xi \rangle x + \Re z}{1 + \langle \xi \rangle}\right)^2 + \frac{\langle \xi \rangle}{1 + \langle \xi \rangle} \left(x - \Re z\right)^2. 
\end{equation}
This and (\ref{eq:defK}) give 
\begin{equation}
\label{eq:mart2}  
 \begin{gathered}
\begin{split}
K ( x, \xi, z ) & = c_n h^{-n/4} \frac{\langle \xi \rangle ^{\frac n 4}}{(1 + \langle \xi \rangle)^{n/2}}
e^{  i \Psi_0 ( x, \xi , z )/h } e^{ -   \Psi_1 ( x, \xi , z ) / h  } e^{ - \Phi ( z ) /h },
\end{split} \\
\Psi_0 ( x, \xi , z ) := x\cdot \xi + \Re z\cdot \Im z - \frac{ (\langle \xi \rangle x + \Re z)\cdot ( \Im z + \xi)  } {1 + \langle \xi \rangle }, \\
\Psi_1 ( x, \xi , z ) := \tfrac 12 \frac{\langle \xi \rangle
( x- \Re z )^2 + ( \xi + \Im z )^2 }  { 1 + \langle \xi \rangle }.
\end{gathered}
\end{equation}}
Suppose now $ ( x, \xi ) \in V $, a complex neighbourhood of $ ( x_0 , \xi_0 ) $. For any $ \varepsilon > 0 $ we can choose $ V $ so that, for some constant $ c_0 > 0$ depending only $ (x_0, \xi_0 ) $, we have for $ ( x, \xi ) \in V$, $ z \in \mathbb C^n$,
\begin{equation}
\label{eq:phases}
 \begin{gathered}  | \Im  \Psi_0 ( x, \xi , z ) | \leq \varepsilon \langle \Im z \rangle \langle \Re z \rangle , \ \   \\
 \Re \Psi_1(x,\xi,z) \geq c_0 \left( ( \Re x - \Re z)^2 + ( \Re \xi + \Im z )^2 \right) - \varepsilon .
 \end{gathered}
 \end{equation}
We now take $ \chi \in C^\infty_{\rm{c}} ( U ; [ 0 , 1 ] ) $ (with $ U $ as in the hypothesis) which is equal to $ 1 $ in $ U_1 = \neigh_{\mathbb C^n }  (  x_0 - i \xi_0) \Subset U $. We then choose $ \varepsilon $ so that with $ \delta $ is given in \eqref{eq:T2T},
 \begin{equation}
 \label{eq:cheps}
  \begin{gathered}
 \varepsilon   \langle \Im z \rangle \langle \Re z \rangle <  \delta/ 4, \ \  z \in U \\
 \tfrac 14 c_0 \left( ( \Re  x - \Re z )^2 + (\Re \xi + \Im z ) ^2 \right) >  \varepsilon
 \langle z \rangle ^2
 , \ \
 z \notin U_1 , \ \ ( x, \xi ) \in V .
 \end{gathered} \end{equation}
 We decompose $ T_{\Lambda_0} u $ as follows
\[   
T_{\Lambda_0} u ( x, \xi ) = A ( x, \xi ) + B ( x, \xi ) , \ \ A ( x, \xi ) := c_n h^{-\frac{3n}2} (T_{\Lambda_0} T_h^* \chi T_h u ) ( x, \xi ). 
\]
From \eqref{eq:mart2}, \eqref{eq:phases}, \eqref{eq:cheps}, and \eqref{eq:T2T} we obtain
\[ \begin{split} | A ( x, \xi ) | &  \leq \mathcal O(1)\, h^{-3n/2- n/4} \int_{\mathbb C^n }
 \chi ( z ) | T_h ( z ) | e^{ 2 \varepsilon \langle \Re z \rangle \langle \Im z \rangle/ h - \Phi ( z ) / h } L( dz )
 \\  &
\leq C_1 e^{ - \delta / 4h } , \ \  ( x, \xi ) \in V ,
\end{split}
\]
where $ C_1 $ depends only on $ ( x_0, \xi_0 ) $.

Next, since $ u \in \mathscr D' ( U_0 ) $ is  $h$-tempered and $ \chi \in C^\infty_{\rm c} ( U_0) $, we have 
\[  \abs{T_h \chi u ( z )} \leq C h^{-N} \langle z \rangle^N  e^{ \Phi ( z ) / h } , \ \ \ z \in \mathbb C^n ,\]
for some $N>0$, see \cite[(1.3.6)]{HiS}.
We use this together with \eqref{eq:mart2}, \eqref{eq:phases}, and \eqref{eq:cheps} to obtain for $(x,\xi) \in V$, 
\[  \abs{B ( x, \xi )} \leq C_2 \int_{ \mathbb C^n \setminus U_1 }
h^{-N} \langle z \rangle^N e^{ |\Im \Psi_0 ( z ) |/h} e^{ - \Re \Psi_1 ( z ) / h } L(dz )
\leq C_3 e^{ - \varepsilon / h } . \]
By summing the estimates for $ A $ and $ B$ we obtain \eqref{eq:T2T2}.
\end{proof}

\begin{proof}[Proof of Proposition \ref{p:global}]
In view of \eqref{eq:WFass} and \eqref{eq:WFa} there exists a (complex) neighbourhood, $V $,  of
$ q^{-1} ( 0 ) \cap \pi^{-1} ( x_0 ) \Subset T^* \mathbb R^n $ such that \eqref{eq:T2T2} holds.

Let $ \Gamma_1  \subset T^* \mathbb R^n $ be an open conic (near infinity) neighbourhood of $ \pi^{-1} ( x_0 ) $ such that
$ \Gamma_1  \cap q^{-1} ( 0 ) \subset V  $ and let $ \Gamma_2 \subset \Gamma_1 $ be
another open conic neighbourhood of $ \pi^{-1} ( x_0 ) $ such that $ \Gamma_2 \cap S^* \mathbb R^n \Subset \Gamma_1 \cap S^* \mathbb R^n$. Choose
$ \psi \in S^0 ( T^* \mathbb R^n ) $ satisfying $ \psi |_{\Gamma_2 } = 1$, $
\supp \psi \subset \Gamma_1 $.
We then choose $ G \in S^1 ( T^* \mathbb R^n ) $, $ \supp G \subset \Gamma_2 $
satisfying \cite[(2.4)]{gaz1} and
\begin{equation}
\label{eq:condG}
| q ( x - i G'_\xi ( x, \xi ), \xi + i G'_x ( x, \xi ) | \geq C \langle \xi \rangle^m , \ \
( x , \xi ) \in \Gamma_1 \setminus V .
\end{equation}
We also put $ G_\varepsilon ( x, \xi ) := \chi_0 ( \varepsilon \xi ) G ( x, \xi ) $, where
$ \chi_0 \in C^\infty_{\rm{c}} ( \mathbb R^n ; [ 0, 1 ] ) $, $ \chi_0  (\xi ) \equiv 1 $ when $|\xi | \leq 1 $.

As in \cite[(2.5)]{gaz1} we then define
\[ \begin{gathered} T_{\Lambda_{G_\varepsilon} }  u ( x, \xi ) := T_{\Lambda_0}  u ( x - i \partial_\xi G_\varepsilon   ( x, \xi ) , \xi + i \partial_xG_\varepsilon   ( x, \xi ) ) , \ \ \
u \in \mathscr S' ( \mathbb R^n ) .
\end{gathered} \]
We note here that since $  G_{\varepsilon} $ is compactly supported, the space defined in
\cite[(2.6)]{gaz2} satisfies $ H_{\Lambda_{G_\varepsilon} }^s = H_h^s ( \mathbb R^n ) $ as a set, but the norm on it
is dramatically different.

We now use \cite[Proposition 2.2]{gaz1}  (with $ h^m P ( x, D_x)  $ replaced by $ \chi P ( x, h D ) $, with $P(x,hD)$ 
given by \eqref{eq:defPP}, with no changes in the proof), to see that, for $ \chi_1 \in C^\infty_{\rm{c}} (
U_0 ) $ equal to one on the support of $\chi $, we have
\begin{equation}
\label{eq:0} \begin{split}  0 & = \| T_{\Lambda_{G_\varepsilon } }  \chi P ( x, h D) \chi_1 u \|^2_{L^2_{\Lambda_{G_\varepsilon}}} \\
& =
\langle \Pi_{\Lambda_{G_\varepsilon} } \bar b_\varepsilon \Pi_{\Lambda_{G_\varepsilon } }
\Pi_{\Lambda_{G_\varepsilon}} b_\varepsilon \Pi_{\Lambda_{G_\varepsilon } }  T_{\Lambda_{G_\varepsilon } }
\chi_1 u , T _{\Lambda_{G_\varepsilon } } \chi_1 u  \rangle_{L^2_{\Lambda_{G_\varepsilon}}}
+ \mathcal O ( h^\infty ) \| \chi_1 u \|^2_{ H_{\Lambda_{G_\varepsilon} }}
, \end{split}
\end{equation}
where
\[ b_{\varepsilon } ( x, \xi ) = q|_{\Lambda_{G_\varepsilon } } =
 q ( x - i \partial_\xi G_\varepsilon   ( x, \xi ) , \xi + i \partial_xG_\varepsilon   ( x, \xi ) )
+  \mathcal O ( h )_{ S^{m} ( \Lambda_{G_\varepsilon } ) }. \]
We note that
\begin{equation}
\label{eq:lowb}    \psi ( x, \xi ) | q|_{\Lambda_{G_\varepsilon}}   ( x, \xi )|^2 \geq  \psi ( x, \xi ) \langle \xi \rangle^{2m}/ C, \ \
( x, \xi ) \notin V .
\end{equation}
We now use \cite[Proposition 6.3]{gaz2} to see that
\[  \Pi_{\Lambda_{G_\varepsilon} } \bar b_\varepsilon \Pi_{\Lambda_{G_\varepsilon } }
\Pi_{\Lambda_{G_\varepsilon}} b_\varepsilon \Pi_{\Lambda_{G_\varepsilon } }  =
\Pi_{\Lambda_{G_\varepsilon} }  | q|_{\Lambda_{G_\varepsilon }} |^2 \Pi_{\Lambda_{G_\varepsilon} }
+ \mathcal O ( h )_{ \langle \xi \rangle^{-2m} L^2_{\Lambda_{G_\varepsilon} } \to
L^2_{\Lambda_{G_\varepsilon} }}  . \]

From the fact that $ u $ is $ h$-tempered we know that for some $N $,
\begin{equation}
\label{eq:someN}
  \| \chi_1 u \|_{H^{-N}_h }^2  \leq C h^{ -2N } .
\end{equation}
Returning to \eqref{eq:0} we see that
\[ \begin{split} &  \langle \psi  | q|_{\Lambda_{G_\varepsilon}}^2
\langle \Re \alpha_\xi \rangle^{-2 N-m}
T_{\Lambda_{G_\varepsilon } }\chi_1 u
, T_{\Lambda_{G_\varepsilon } }\chi_1 u \rangle_{ L^2_{\Lambda_{G_\varepsilon } } } \\
& \ \ \ \  \leq
C \langle ( 1 - \psi ) \langle \xi \rangle^{- 2 N} T_{\Lambda_{G_\varepsilon } }\chi_1 u,
T_{\Lambda_{G_\varepsilon } }\chi_1 u \rangle_{L^2_{\Lambda_{G_\varepsilon } } }
+ \mathcal O ( h ) \| \langle \Re \alpha_\xi \rangle^{-N}  T_{\Lambda_{G_\varepsilon } }\chi_1 u \|_{L^2_{\Lambda_{G_\varepsilon } }}^2.
\end{split} \]
The left hand side can be bounded from below using \eqref{eq:lowb} and in the first term on the
right hand side $ \Lambda_{G_\varepsilon } $ can be replaced by $ \Lambda_0 $ since
$ G \equiv 0$ near the support of $  1 - \psi $. We can then bound that term by
\[  \| u \|^2_{ H^{-N}_{\Lambda_0 }} \simeq \| \chi_1 u \|_{ H^{-N}_h }^2.\]
Hence we obtain (using the definition of the
norm on $ L^2_{\Lambda_{G_\varepsilon } } $ in \cite[(2.6)]{gaz1})
\begin{equation}
\label{eq:Hn2Hm} \| \chi_1 u\|^2_{ H^{-N}_{\Lambda_{G_\epsilon } } }  \leq C \| \chi_1 u \|_{H^{-N}_h } ^2
+ C  \int_{ \Lambda_{ G_{\varepsilon } } \cap V }
 | T_{\Lambda_{G_\varepsilon } } \chi_1 u  ( \alpha )|^2
e^{ - 2 H_\varepsilon ( \alpha ) /h } d \alpha .
\end{equation}
Lemma \ref{l:T2T} shows $ | T_{\Lambda_{G_\varepsilon } } \chi_1 u  ( \alpha )|^2 |
= \mathcal O ( e^{ - \delta'/h } ) $ for $ \alpha \in \Lambda_{G_\varepsilon } \cap V $.
Hence, if $ \varepsilon_0 $ in \cite[(2.4)]{gaz1} is chosen sufficiently
small (so that $ H_\varepsilon $ is small near $ V $), the last term on the right hand side is bounded
uniformly in $ \varepsilon $ and $ h$.  Using \eqref{eq:someN} we obtain
\begin{equation}
\label{eq:someNN}
 \| \chi_1 u\|^2_{ H^{-N}_{\Lambda_{G_\varepsilon } } }   \leq C h^{-2N}
 \end{equation}
with $ C$ independent of $ \varepsilon $.

We now choose yet another conic (near infinity) neighbourhood of $ \pi^{-1} (x_0 ) $,
$ \Gamma_3 \subset \Gamma_2 $ and such that $ G ( x, \xi ) = \varepsilon_0  \langle \xi \rangle/2
$ in $ \Gamma_3 $ (note that $ G $ is supported in $ \Gamma_2 $).
Since $ G_\varepsilon ( x, \xi ) = \chi_0 ( \varepsilon \xi ) G ( x, \xi )$,
and the estimates are uniform in $ \varepsilon $, the monotone convergence theorem,
\eqref{eq:someNN} and  \cite[Proposition 2.5]{gaz1} give
\[   \int_{\Gamma_3 } e^{ \varepsilon_0 \langle \xi \rangle/2 h } | T_{\Lambda_0 } \chi_1 u ( x, \xi ) |^2
dx d \xi  \leq e^{-\varepsilon_0 / 6 h } . \]
(We first obtain a bound $ \mathcal O ( h^{-2N} ) $ for a weight $ e^{2\varepsilon_0 \langle \xi \rangle/3} $
which then gives the bound above.)

The inversion formula \cite[(2.2)]{gaz1}, \cite[Proposition 2.2]{gaz2} easily shows that
$ u $ can be holomorphically continued to a neighbourhood of $ x_0 $ in $ \mathbb C^n $
and that it is bounded by $\mathcal O(1) e^{-\delta / h } $. We then get \eqref{eq:expdec1}  from Cauchy
estimates.
\end{proof}

\section*{Appendix by Zhongkai Tao and Maciej Zworski}
\renewcommand{\theequation}{A.\arabic{equation}}
\refstepcounter{section}
\renewcommand{\thesection}{A}
\setcounter{equation}{0}

In this appendix we show how to solve the eikonal equation \eqref{eq:eik1} at the
corners of the hexagon and prove that \eqref{eq:kash} holds for the corresponding
weight $ \Phi$. Once that is done the proof of exponential decay
 proceeds the same as that of Theorem \ref{t:3} (which applied to points in the interior
 of the edges) combined with global elliptic estimates of Proposition \ref{p:global}.
However, there is no need for a preparatory canonical transformation of \S \ref{s:prince}
and the analysis of \S \ref{s:eik} is replaced by this appendix.

We start by recalling \eqref{eq:Tayq}:
\[ p ( z, \zeta ):= q ( z_S + z , \zeta ) = 4  \bar \zeta^2  + i A  \bar z - B z^2 + \mathcal O ( |z|^3 ) , \ \ \
A, B > 0. \]
Making a symplectic change of variables $  ( w, \zeta_1 ) \mapsto ( z, \zeta ) $,
$ z = -  i \alpha w $ and $ \zeta =  i \alpha^{-1} \zeta_1 $, $ \alpha =  A/B $,
we obtain
\[ \begin{split}
p ( z , \zeta ) & = p_1 ( w , \zeta_1 ) = - 4 \alpha^{-2} \bar \zeta_1^2 - \alpha A \bar w + B \alpha^2 w^2 +
\mathcal O ( |w|^3 ) \\
& =  -A^2 B^{-1} (4 B^3 A^{-4}  \bar \zeta_1^2 + \bar w - w^2 +  \mathcal O ( |w|^3 )  ) .
\end{split} \]
Hence, by a change of variables and by rescaling the semiclasical parameter $ h $ by the fixed constant $ 4 B^3 A^{-4} $, we can assume that
the stacking point is given by $ w = 0 $ and that the principal symbol is given by
\begin{equation}
\label{eq:p0} p ( w, \zeta ) = p_0 ( w, \zeta ) + \mathcal O ( |w|^3 ) , \ \ \
p_0 ( w, \zeta ) := \bar \zeta^2 + \bar w - w^2 . \end{equation}

We consider the eikonal equation \eqref{eq:eik1} for
\eqref{eq:p0}: with $ z \in \mathbb C^2 $ and $ w, v \in \mathbb C $,
\begin{equation}
\label{eq:eik2}   \partial_{z_1} \varphi_0 ( z, w , v ) = ( \partial_{v  } \varphi_0 ( z, w , v) )^2 +
v - w^2 +\mathcal{O}(|(v,w)|^3),   \end{equation}
$ \varphi ( 0 , z_2 , w, v  ) = i w v + w z_2 $.
We note that $ v = \bar w $ corresponds to the real $ y \in \mathbb R^2 $ in \eqref{eq:eik1}.
The boundary condition guarantees that $ \varphi $ satisfies \eqref{eq52} with
$ z = x \in \mathbb C^2 $, $ x_0 = 0 $,  $ w = y_1 + i y_2 $, $ y \in \mathbb R^2 $, $ ( y_0, \eta_0 ) = 0 $
(corresponding to the characteristic point $ ( 0, 0 ) $ for $ p_0 $), $v =  \bar w = y_1 - i y_2 $;
we check that the equation and the initial condition give at $ z = ( w, v ) = ( 0 , 0 ) $,
\[   \begin{pmatrix}  \partial^2_{z_1 w } \varphi  & \partial^2_{ z_1  v } \varphi  \\
\partial^2_{z_2 w } \varphi & \partial^2_{ z_2  v } \varphi  \end{pmatrix} =
\begin{pmatrix} 0 & 1 \\ 1 &  0 \end{pmatrix}, \ \ \
\Im \partial_y^2 [ \varphi |_{ w = y_1 + iy_2, v = \bar w } ] =  2 I_{\mathbb R^2} . \]

As for \eqref{eq111}, the solution exists by the Cauchy--Kowalevski Theorem, for $ z \in \neigh_{\mathbb C^2 } ( 0 ) $ and
$ y \in \neigh_{\mathbb C^2 } ( 0 ) $. We then have
\begin{equation*}
\begin{aligned}
    \varphi(z_1,z_2,w,\Bar{w}) 
    &=iw\Bar{w}+z_2w +\sum_{j=1}^6\frac{z_1^j}{j!} \partial_{z_1}^j\varphi(0,z_2,w,\Bar{w})+\mathcal{O}(|z_1|^7).
\end{aligned}
\end{equation*}
We compute each term:
    \begin{align*}
        \partial_{z_1}\varphi(0,z_2,w,\bar{w})=\bar{w}-2w^2+\mathcal{O}(|w|^3).
    \end{align*}\begin{align*}\partial_{z_1}^2\varphi(0,z_2,w,\Bar{w}) & =\partial_{z_1}(\partial_{\Bar{w}}\varphi)^2=2\partial^2_{\Bar{w}z_1}\varphi\partial_{\Bar{w}}\varphi
     =2(1+\mathcal{O}(|w|^2))(iw) \\ & =2iw+\mathcal{O}(|w|^3) ,
    \end{align*}
    \begin{align*}
        \partial_{z_1}^3\varphi(0,z_2,w,\Bar{w})&=\partial^2_{z_1}(\partial_{\Bar{w}}\varphi)^2=2\partial_{\Bar{w}}\varphi\partial^3_{\Bar{w}z_1z_1}\varphi+2(\partial^2_{z_1\Bar{w}}\varphi)^2\\
        &=2iw\partial^2_{\Bar{w}z_1}(\partial_{\Bar{w}}\varphi)^2+2(1+\mathcal{O}(|w|^2))^2\\
        &=2+4iw\partial_{\Bar{w}}\varphi\partial^3_{\Bar{w}\Bar{w}z_1}\varphi+\mathcal{O}(|w|^2) = 2+\mathcal{O}(|w|^2),
    \end{align*}
\begin{align*}
        \partial_{z_1}^4\varphi(0,z_2,w,\Bar{w})&=\partial_{z_1}^3(\partial_{\Bar{w}}\varphi)^2=2\partial_{\Bar{w}}\varphi\partial^4_{z_1z_1z_1\Bar{w}}\varphi+6\partial^2_{\Bar{w}z_1}\varphi\partial^3_{z_1z_1\Bar{w}}\varphi\\
        &=2iw\partial^3_{z_1z_1\Bar{w}}(\partial_{\Bar{w}}\varphi)^2+6(1+\mathcal{O}(|w|^2))\partial^2_{z_1\Bar{w}}(\partial_{\Bar{w}}\varphi)^2\\
        &=2iw(2\partial^4_{z_1z_1\Bar{w}\Bar{w}}\varphi\partial_{\Bar{w}}\varphi+4\partial^3_{z_1\Bar{w}\Bar{w}}\varphi\partial^2_{z_1\Bar{w}}\varphi)+\mathcal{O}(1)\partial^3_{z_1\Bar{w}\Bar{w}}\varphi\partial_{\Bar{w}}\varphi\\
        &=\mathcal{O}(|w|)\partial^3_{z_1\Bar{w}\Bar{w}}\varphi+\mathcal{O}(|w|^2)\\
        &=\mathcal{O}(|w|)\partial^2_{\Bar{w}\Bar{w}}((\partial_{\Bar{w}}\varphi)^2+\Bar{w}-w^2+\mathcal{O}(|w|^3))+\mathcal{O}(|w|^2) =\mathcal{O}(|w|^2),
    \end{align*}
 \begin{align*}
        \partial_{z_1}^5\varphi(0,z_2,w,\Bar{w})&=\partial_{z_1}^4(\partial_{\Bar{w}}\varphi)^2=2\partial_{\Bar{w}}\varphi\partial^5_{z_1z_1z_1z_1\Bar{w}}\varphi+8\partial^2_{z_1\Bar{w}}\varphi\partial^4_{z_1z_1z_1\Bar{w}}\varphi+6(\partial^3_{z_1z_1\Bar{w}}\varphi)^2\\
        &=\mathcal{O}(|w|) +\mathcal{O}(|w|^4) =\mathcal{O}(|w|),
    \end{align*}
    \begin{align*}
        \partial_{z_1}^6\varphi(0,z_2,w,\Bar{w})&=\partial_{z_1}^5(\partial_{\Bar{w}}\varphi)^2
        =2\partial_{\Bar{w}}\varphi\partial^6_{z_1z_1z_1z_1z_1\Bar{w}}\varphi+10\partial^2_{z_1\Bar{w}}\varphi\partial^5_{z_1z_1z_1z_1\Bar{w}}\varphi+20\partial^3_{z_1z_1\Bar{w}}\varphi\partial^4_{z_1z_1z_1\Bar{w}}\varphi\\
        &=\mathcal{O}(|w|)+\mathcal{O}(1)\partial^4_{z_1z_1z_1\Bar{w}}(\partial_{\Bar{w}}\varphi)^2+\mathcal{O}(|w|^3)\\
        &=\mathcal{O}(1)\partial^5_{z_1z_1z_1\Bar{w}\Bar{w}}\varphi\partial_{\Bar{w}}\varphi+\mathcal{O}(1)\partial^4_{z_1z_1\Bar{w}\Bar{w}}\varphi\partial^2_{z_1\Bar{w}}\varphi+\mathcal{O}(1)\partial^3_{z_1\Bar{w}\Bar{w}}\varphi\partial^3_{z_1z_1\Bar{w}}\varphi+\mathcal{O}(|w|)\\
        &=\mathcal{O}(1)\partial^4_{z_1z_1\Bar{w}\Bar{w}}\varphi+\mathcal{O}(|w|)
        =\mathcal{O}(1)\partial^3_{z_1\Bar{w}\Bar{w}}(\partial_{\Bar{w}}\varphi)^2+\mathcal{O}(|w|)\\
        &=\mathcal{O}(1)\partial^4_{z_1\Bar{w}\Bar{w}\Bar{w}}\varphi\partial_{\Bar{w}}\varphi+\mathcal{O}(|w|)=\mathcal{O}(|w|).
    \end{align*}
In conclusion, we have
\begin{equation*}
    \varphi=iw\Bar{w}+z_2w+z_1(\Bar{w}-2w^2)+iz_1^2w+\tfrac{1}{3}z_1^3+\mathcal{O}(|z_1w^3|+|z_1^3w^2|+|z_1^5w|+|z_1|^7).
\end{equation*}
To obtain  $\Phi ( z ) := \sup_{ |w| < \varepsilon , w \in \mathbb C} - \Im \varphi ( z , w , \bar w ) $,
we find the critical point, $  w  =w(z)$,   given by solving
\begin{equation}\label{e:corner1}
    0=\partial_w\varphi-\overline{\partial_{\Bar{w}}\varphi}=2i\Bar{w}+z_2-\Bar{z}_1-4wz_1+iz_1^2+\mathcal{O}(|z_1w^2|+|z_1^3w|+|z_1|^5),
   \end{equation}
 which, by analytic implicit function theorem, gives an analytic solution $w=w(z)$ near $z=0$:
\begin{equation}
\label{eq:2bw} w ( z ) = \tfrac{1}{2} i ( z_1 - \bar z_2 ) + \mathcal O ( |z|^2 ) .
\end{equation}
We then find
$$ \Psi ( z_2 ) : = \inf_{ |z_1 | < \varepsilon } \Phi ( z ) ,$$
by looking for the critical point, $z_1=z_1(z_2)$,  solving
\begin{equation}\label{e:corner2}
    0=2i\partial_{z_1}\Im \varphi=\Bar{w}-2w^2+2iz_1w+z_1^2+\mathcal{O}(|w^3|+|z_1^2w^2|+|z_1^4w|+|z_1|^6),
\end{equation}
where $w =  w (z) $ is given by \eqref{eq:2bw}. In view of \eqref{eq:2bw} the solution,  $ z_1 = z_1 ( z_2 ) $, satisfies $ z_1 = \bar z_2 + \mathcal O ( |z_2|^2 ) $ and we have
$w( z_1 (z_2) , z_2 ) =\mathcal{O}(|z_2|^2)$. Hence we can solve \eqref{e:corner2} up to fifth order terms. Inserting \eqref{e:corner1} into the first term of \eqref{e:corner2}, we get (with $ w = w ( z ) $)
\begin{equation*}
    -\tfrac{1}{2i}(z_2-\Bar{z}_1-4w z_1+iz_1^2)-2w ^2+2iz_1w+z_1^2=\mathcal{O}(|z_2|^5),
\end{equation*}
or
$   \frac{1}{2i}(z_2-\bar{z}_1)=\frac{1}{2}z_1^2-2w^2+\mathcal{O}(|z_2|^5)$.
It follows that 
\begin{equation*}
    z_1=\Bar{z}_2+iz_2^2+2\Bar{z}_2^2z_2+\mathcal{O}(|z_2|^4),\quad {w} (z_1(z_2), z_2) =-{z}_2^2+\mathcal{O}(|z_2|^4).
\end{equation*}
We can then compute the fourth order term. However, we observe that $\mathcal{O}(|z_2|^4)$ terms affect the fifth order term of $\varphi$ only in $z_2w+z_1\Bar{w}$, and do not affect the imaginary part of it, so we conclude
\begin{equation*}
    \Psi(z_2)=\tfrac{1}{3}\Im(z_2^3)+|z_2|^2\Im(z_2^3)+\mathcal{O}(|z_2|^6).
\end{equation*}

The first term in the expansion of $ \Psi $ is harmonic but the second,  $|\zeta|^2\Im(\zeta^3)$,  is not subharmonic. For any subharmonic function $u(\zeta)\leq |\zeta|^2\Im(\zeta^3)$,
$ |\zeta| \leq 1 $, we have
\begin{equation*}
    u(0)\leq \pi^{-1}\int_{|\zeta|\leq 1}u(\zeta)\,L(d\zeta)<\pi ^{-1}\int_{|\zeta|\leq 1}|\zeta|^2\Im(\zeta^3)\,L(d\zeta)=0.
\end{equation*}
Taking $u$ to be the subharmonic minorant of $|\zeta|^2\Im(\zeta^3)$ in the unit disk (see Lemma \ref{l:minorant}
and its proof) we conclude that
\begin{equation}
\label{eq:min2}
\begin{split}
\exists \, c_0>0, \ \ & u(0)\leq -c_0<0, \ \text{for any subharmonic function  in $ \{ | \zeta| \leq 1\}  $}
\\
& \text{\ \ \ \ \ \ \ \ \ satisfying $u(\zeta)\leq |\zeta |^2\Im(\zeta^3)$.}
\end{split}
\end{equation}

Now suppose there is a (continuous) subharmonic function $u$ in $\{|z|\leq \delta\}$ such that $u(z)\leq \Psi(z)$. Then $\Tilde{u}(z)=u(z)-\frac{1}{3}\Im(z^3)$ is also subharmonic and $\Tilde{u}(z)\leq |z|^2\Im(z^3)+\mathcal{O}(|z|^6)$. After rescaling, $\delta^{-5}\Tilde{u}(\delta z)$ is a subharmonic function defined in the unit disk and
\begin{equation*}
    \delta^{-5}\Tilde{u}(\delta z)\leq |z|^2\Im(z^3)+\mathcal{O}(\delta).
\end{equation*}
From \eqref{eq:min2}  we have
\begin{equation*}
    \delta^{-5}u(0)=\delta^{-5}\Tilde{u}(0)\leq -c_0+\mathcal{O}(\delta)\leq -c_0/2<0.
\end{equation*}
for $\delta$ sufficiently small. This gives \eqref{eq:kash} and the proof of exponential decay
proceeds as in \S \ref{s:pr2}.

\end{document}